\documentclass[journal=jacsat,manuscript=article]{achemso}

\usepackage[version=3]{mhchem} % Formula subscripts using \ce{}
\usepackage{physics}
\usepackage{pdfpages}
\author{Rosario R. Riso}
\email{rosario.r.riso@ntnu.no}
\author{Matteo Castagnola}
\author{Enrico Ronca}
\altaffiliation{Dipartimento di Chimica, Biologia e Biotecnologie, Università degli Studi di Perugia,
Via Elce di Sotto, 8,06123, Perugia, Italy}
\author{Henrik Koch}
\affiliation{Department of Chemistry, Norwegian University of Science and Technology, 7491 Trondheim, Norway}
\email{henrik.koch@ntnu.no}

\title[Chiral polaritonics]{Chiral polaritonics: cavity-mediated enantioselective excitation condensation}


\begin{document}

\begin{tocentry}
\includegraphics[width=\textwidth]{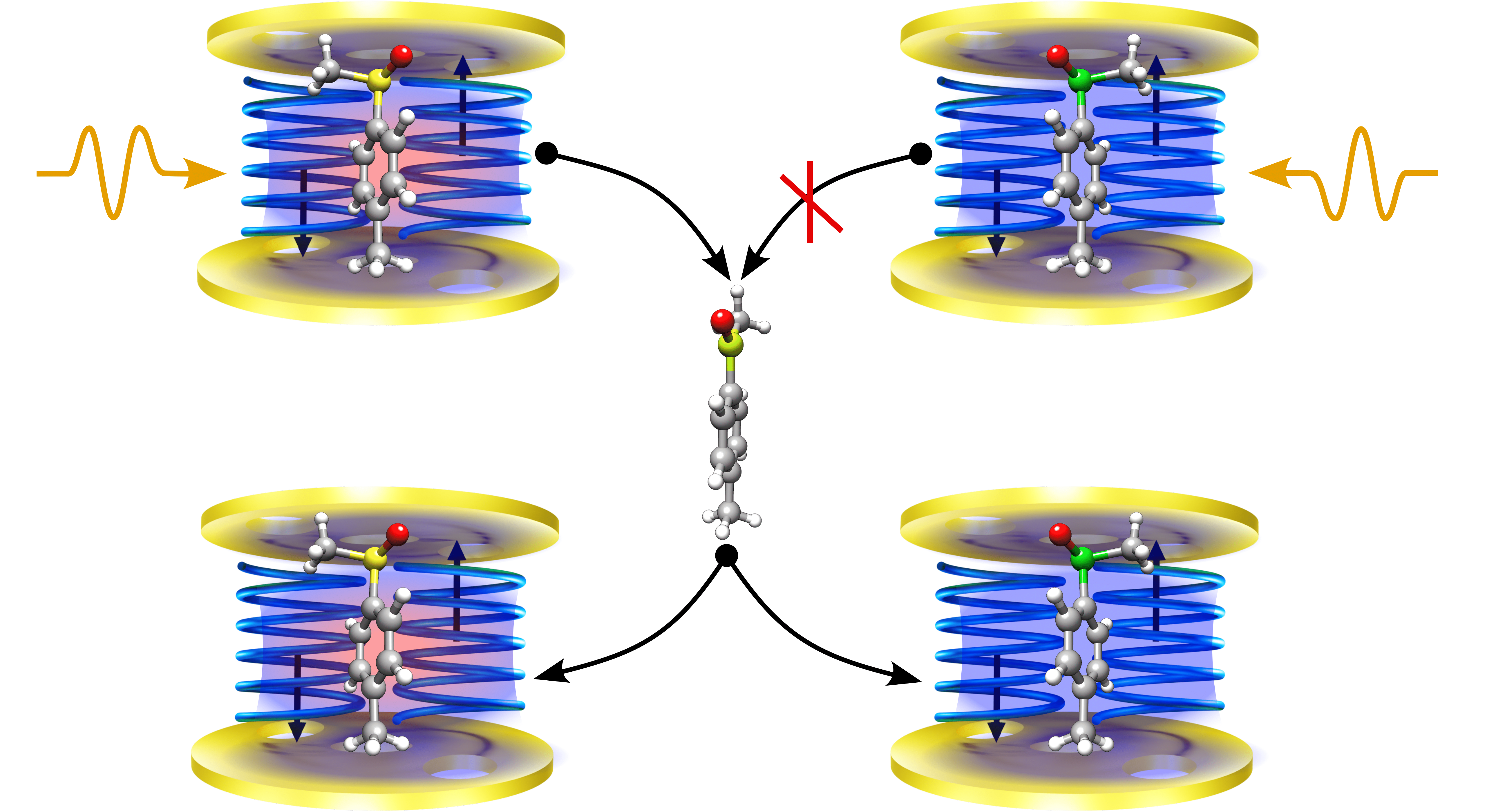} 
\end{tocentry}
\begin{abstract}
Separation of the two mirror images of a chiral molecule, the enantiomers, is a historically complicated problem of major relevance for biological systems. Since chiral molecules are optically active, it has been speculated that strong coupling to circularly polarized fields may be used as a general procedure to unlock enantiospecific reactions. In this work, we focus on how chiral cavities can be used to drive asymmetry in the photochemistry of chiral molecular systems. We first show that strong coupling to circularly polarized fields leads to enantiospecific Rabi splittings, an effect that displays a collective behavior in line with other strong coupling phenomena.
Additionally, entanglement with circularly polarized light generates an asymmetry in the enantiomer population of the polaritons, leading to a condensation of the excitation on a preferred molecular configuration. These results confirm that chiral cavities represent a tantalizing opportunity to drive asymmetric photochemistry in enantiomeric mixtures. 
\end{abstract}
\section{Introduction}
Strong coupling between different parts of a multicomposite system is achieved when the interaction between two components becomes intense enough to entangle them, overcoming dissipation and decoherence processes  \cite{dovzhenko2018light,hameau1999strong,teufel2011circuit}. 
In the area of quantum optics \cite{loudon2000quantum}, this regime can be achieved using a large variety of apparatuses such as optical cavities\cite{garcia2021manipulating,wright2023rovibrational,wright2023versatile,damari2019strong,schwartz2013polariton,thomas2016ground,simpkins2023control}, nanoparticles\cite{baumberg2019extreme,baranov2020ultrastrong}, and circuits\cite{haroche2020cavity,teufel2011circuit,wallraff2004strong}. These devices confine electromagnetic fields in small volumes with the effect of enhancing the strength of light-matter interactions \cite{fitzgerald2016quantum,chikkaraddy2016single}. 
The superposition between photons and matter leads to the emergence of hybrid states known as polaritons \cite{castagnola2024polaritonic,flick2017atoms,martinez2018can,feist2018polaritonic,herrera2020molecular,fregoni2023polaritonic}. The polaritons exhibit unique properties beyond those of their individual electronic and photonic components. 
Moreover, the partial photonic nature of the polaritons allows us to tailor the main features of these states by fine-tuning the specifics of the optical device, i.e. its geometry or the material of the mirrors\cite{mandal2019investigating}. 
Examples of adjustable field features include the frequency of the photons, as well as the shape of the photonic wave.

\begin{figure}
    \centering
    \includegraphics[width=\textwidth]{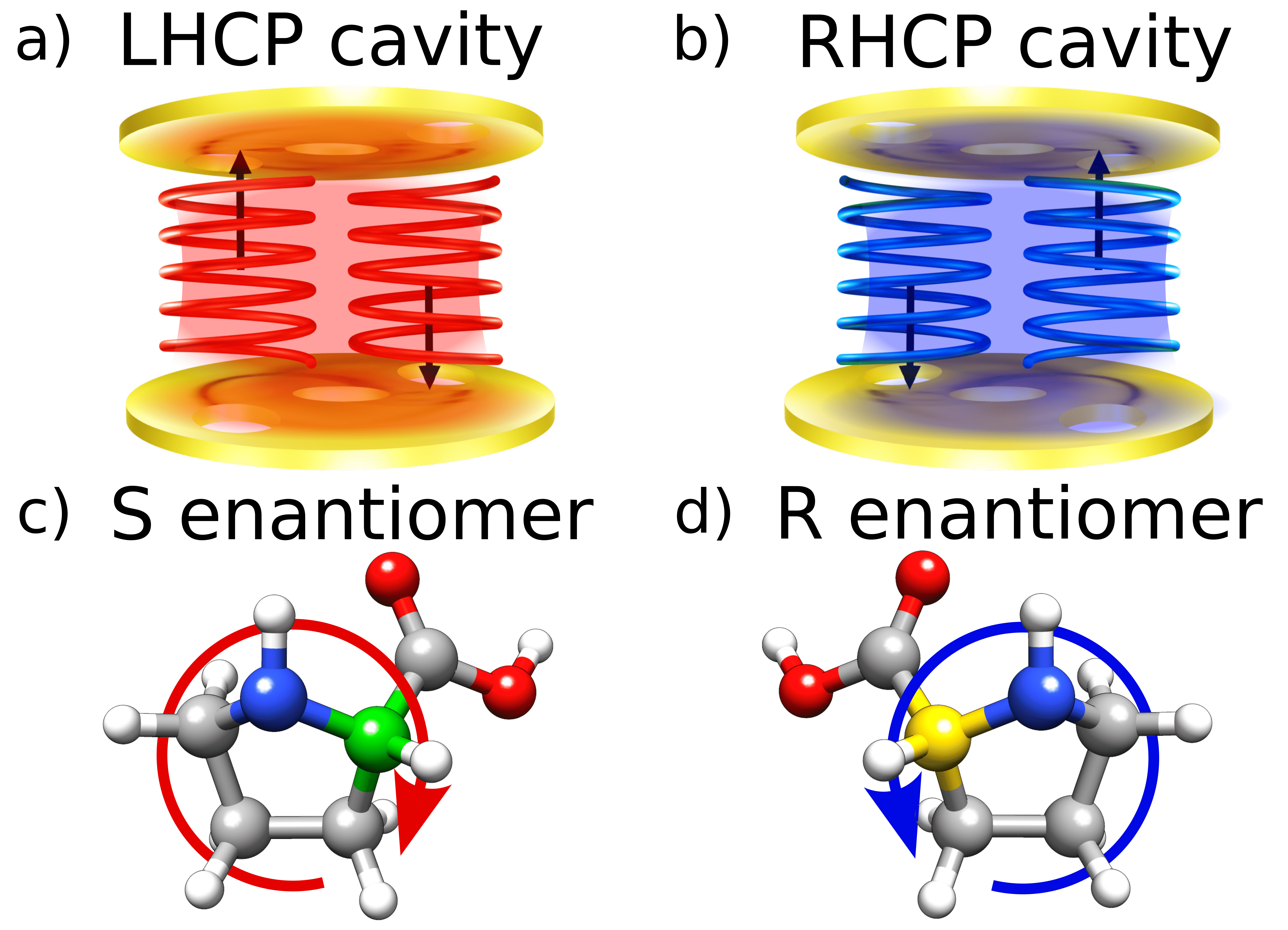}
    \caption{\textbf{Field and molecular chirality.} a)-b) Chiral cavities where only left-handed circularly polarized (LHCP) or right-handed circularly polarized (RHCP) fields are confined. The spirals are a pictorial representation c)-d) S and R enantiomers of proline, one of the amino acids composing proteins in our body. The enantiomeric identification protocol follows three steps: 1) assigning a priority number to each atom attached to the chiral center based on the atomic number (from largest to smallest); 2) Rotating the structure such that the lightest element (H in this case) is pointing backward; 3) Drawing an arrow from the highest to the lowest priority element. If the arrow rotates clockwise, we refer to the enantiomer as \textit{R}, otherwise we refer to the enantiomer as \textit{S}. This procedure provides a unique identification of the molecular structure \cite{clayden2012organic}.
    %Reduce text size
    }
    \label{fig:chirality}
\end{figure}

In this regard, recent experimental works report the fabrication of mirrors that selectively reflect one circular polarization of the field while preserving the field handedness upon reflection\cite{plum2015chiral,liu2020switchable,semnani2020spin,wu2023bottom}. Opportune combinations of such mirrors allows us to produce chiral cavities, see Fig.\ref{fig:chirality}a/b \cite{taradin2021chiral,gautier2022planar,sun2022polariton,voronin2022single}. Since circularly polarized fields can differentiate between the two forms of a chiral molecule, the enantiomers, it is reasonable to expect that the energy degeneracy normally observed between the enantiomers should be lifted when strongly coupled with circularly polarized fields\cite{PhysRevA.107.L021501,sun2022polariton,baranov2022towards,schafer2023chiral,ke2023vacuum}, see Fig.\ref{fig:chirality}c/d.  
Theoretical modeling of the strong light-matter coupling regime offers critical insight into the principles governing the complex interplay between light and matter, allowing us to predict qualitative phenomena before performing actual experiments \cite{liebenthal2022equation,vu2022enhanced,gudem2023cavity}. Recently, we 
demonstrated theoretically that, for the ground state, strong coupling to circularly polarized fields induces energy differences between the enantiomers and that such effects can in principle be used to generate an asymmetry in otherwise fully symmetric synthetic processes \cite{riso2023strong,riso2023strongg}. The predicted effects do however only become measurable for light-matter couplings beyond current experimental capabilities. Much larger field-induced modifications can instead be achieved for excited state properties, with observable effects at experimentally realizable quantization volumes. In the case of Fabry-Pèrot cavities, for example, notable effects on molecular excited states include delayed emission, modification of excitation energies\cite{ebbesen2023introduction,berghuis2019enhanced} as well as modification of photochemical \cite{hutchison2012modifying,schwartz2024importance} and thermal reactivity \cite{hirai2023molecular,nagarajan2021chemistry}. Specifically, the results presented by Ebbesen and coworkers \cite{thomas2019tilting,lather2019cavity,thomas2020ground} and by Simpkins and coworkers \cite{ahn2023modification} clearly show that placing the field in resonance with vibrational excitations of a molecule is a viable approach to slow down and catalyze reactions. Similar effects should also be observable in chiral cavities, with the key difference that the field effects should be enantiospecific \cite{baranov2022towards,schafer2023chiral,bustamante2024relevance}.  
Strong coupling between molecules and circularly polarized fields in a chiral cavity has not been experimentally realized yet. However, investigations of these systems have been presented in recent theoretical works, either based on a dielectric description of the molecular medium or using a Jaynes-Cummings approach \cite{mauro2022charge,schafer2023chiral,baranov2022towards}. The present work complements these previous results as we provide an \textit{ab initio} description of excited states in chiral cavities and advance the idea that strong coupling to circularly polarized fields can be used to differentiate between the photochemistry of chiral molecules.
The paper is structured as follows: we first present the theoretical framework to describe strong light-matter coupling in chiral cavities discussing the symmetry properties of the Hamiltonian. We further illustrate the key steps in the development of our \textit{ab initio} methodology for the calculation of excited states in a QED environment. In the Results section, we show that due to the chirality of the field, different enantiomers of the same molecule exhibit different Rabi splittings in the cavity. Moreover, we illustrate that the differences in the observed Rabi splittings display the same collective behavior observed in the case of linearly polarized fields \cite{castagnola2024collective,sidler2023unraveling}. This mechanism allows us to increase the chirality effects, up to the point where the photochemistry of the two enantiomers could be significantly different.  
\section{Chiral cavity Hamiltonian and its symmetries}
To describe the interaction of a molecule with a chiral cavity we use the minimal coupling Hamiltonian\cite{ruggenthaler2023understanding,svendsen2023theory} 
\begin{equation}
\begin{split}
H &=\sum_{i}\frac{1}{2m_{i}}\left(\mathbf{p}_{i}+Z_{i}\mathbf{A}(\mathbf{r}_{i})\right)^{2}+ \sum_{I}\frac{1}{2M_{I}}\left(\mathbf{p}_{I}-Z_{I}\mathbf{A}(\mathbf{R}_{I})\right)^{2}\\
&-\frac{1}{4\pi\epsilon_{0}}\sum_{i, I}\frac{Z_{i}Z_{I}}{\left|R_{I}-r_{i}\right|}+\frac{1}{4\pi\epsilon_{0}}\sum_{i>j}\frac{Z_{i}Z_{j}}{\left|r_{i}-r_{j}\right|}+\frac{1}{4\pi\epsilon_{0}}\sum_{I>J}\frac{Z_{I}Z_{J}}{\left|R_{I}-R_{J}\right|}  \label{eq:Minimal}\\
&+\frac{\epsilon_{0}}{2}\int\left(\mathbf{E}^{2}(\mathbf{r})+c^{2}\mathbf{B}^{2}(\mathbf{r})\right)d^{3}r,
\end{split}    
\end{equation}
where the electronic ($i$ and $j$) and nuclear ($I$ and $J$) momenta $\mathbf{p}$ are shifted by the cavity vector potential $\mathbf{A}(\mathbf{r})$ evaluated at the electronic and nuclear positions, $\mathbf{r}_{i}$ and $\mathbf{R}_{I}$, multiplied with the charge of the particle, $Z_{i}$ and $Z_{I}$. The electric and the magnetic fields are denoted by $\mathbf{E}(\mathbf{r})$ and $\mathbf{B}(\mathbf{r})$, respectively. In Equation \ref{eq:Minimal}, $m_{i}$ and $M_{I}$ denote the electronic and nuclear masses while $\epsilon_{0}$ is the vacuum dielectric constant. It is critical to account for the quantum nature of the electromagnetic field to accurately describe the cavity-induced effects in the strong coupling regime. Therefore, the field is modeled employing a quantum electrodynamics treatment \cite{weight2023investigating,foley2023ab,ruggenthaler2023understanding}. 
The field features are encoded in the functional shape of the vector potential, which, in a chiral cavity, reads\cite{riso2023strong} 
\begin{equation}
\mathbf{A}_{\pm}(\mathbf{r}) =\sum_{k} \frac{\lambda}{\sqrt{2\omega_{k}}}\left(\boldsymbol{\epsilon}_{k\pm}b_{k}e^{i\mathbf{k}\mathbf{r}}+\boldsymbol{\epsilon}^{*}_{k\pm}b^{\dagger}_{k}e^{-i\mathbf{k}\mathbf{r}}\right).
\label{eq:Shape}
\end{equation}
In Equation \ref{eq:Shape}, the photonic operators $b_{k}$ and $b^{\dagger}_{k}$ respectively annihilate and create a photon with wave vector $k$ and frequency $\omega_{k}$ while $\boldsymbol{\epsilon}_{k\pm}$ denotes the field polarization. The parameter $\lambda$ quantifies the strength of the light-matter coupling, i.e. when $\lambda=0$ a.u. there is no coupling between the photons and the molecule. Conversely, increasing values of $\lambda$ correspond to an enhancement of the field effects on the molecule. The square of this parameter is inversely proportional to the quantization volume of the field, namely $\lambda=\sqrt{\frac{\hbar}{\epsilon_{0}V}}$. 
Experimentally achievable values for $\lambda$ in Fabry-Pèrot cavities range up until $\lambda\approx0.05$ a.u., corresponding to a quantization volume of around 1 nm$^{3}$\;\cite{chikkaraddy2016single,santhosh2016vacuum}. 
In the field expansion in Eq.\ref{eq:Shape}, we notice that an exact treatment of the field requires the inclusion of all the cavity-allowed modes \cite{svendsen2023theory}. However, using the full vector potential is computationally unfeasible. Moreover, the largest field-induced effects are observed when one of the cavity photons is resonant or closely resonant with a molecular excitation. In the case where one single mode of the field is resonant with a molecular excitation, the degenerate molecular and photonic states give rise to two polaritonic states, the upper and lower polaritons, where both excitations are significantly featured. In a simplified scheme, the wave function for such states is
\begin{eqnarray}
&\ket{\psi}_{LP}=\ket{E,0}C_{E}+\ket{G,1}C_{G} \\
&\ket{\psi}_{UP}=\ket{E,0}C_{G}-\ket{G,1}C_{E} \\
&C^{2}_{E}+C^{2}_{G}=1, 
\end{eqnarray}
with $\ket{E,0}$ being the state where the molecule is excited and there are no photons in the cavity, while $\ket{G,1}$ labels the state where the molecule is in its ground state and the cavity photon is excited. To describe the main field effects, therefore, it is sufficient to only include the resonant photons in the field expansion. Since in our study we are assuming the mirrors to be perfect reflectors for a selected circular polarization of the field, both the modes with $+k$ and -$k$ must be included in Eq.\ref{eq:Shape}.
Assuming that the field polarization is
\begin{equation}
\boldsymbol{\epsilon}_{k\pm}=\boldsymbol{\epsilon}_{\pm}=\frac{1}{\sqrt{2}}\left(
1, \;\; \pm i , \;\; 0\right),    
\end{equation}
the two-mode Hamiltonian reads 
\begin{equation}
\begin{split}
H_{\pm} =&\sum_{i}\frac{p^{2}_{i}}{2}+\sum_{I}\frac{p^{2}_{I}}{2M_{I}}-\sum_{i, I}\frac{Z_{I}}{\left|r_{i}-R_{I}\right|}+\sum_{i>j}\frac{1}{\left|r_{i}-r_{j}\right|}+\sum_{I>J}\frac{Z_{I}Z_{J}}{\left|R_{I}-R_{J}\right|}\\
-&\frac{\lambda}{\sqrt{2\omega_{k}}}\sum_{i}\left[\left(\mathbf{p}_{i}\cdot\boldsymbol{\epsilon}_{\pm}\right)e^{i\mathbf{k}\mathbf{r}_{i}}\left(b_{k}+b^{\dagger}_{-k}\right)
+\left(\mathbf{p}_{i}\cdot\boldsymbol{\epsilon}^{*}_{\pm}\right)e^{-i\mathbf{k}\mathbf{r}_{i}}\left(b^{\dagger}_{k}+b_{-k}\right)\right]\label{eq:Multimode}
\\  
+&\frac{\lambda}{\sqrt{2\omega_{k}}}\sum_{I}\left[\frac{Z_{I}}{M_{I}}\left(\mathbf{p}_{I}\cdot\boldsymbol{\epsilon}_{\pm}\right)e^{i\mathbf{k}\mathbf{R}_{I}}\left(b_{k}+b^{\dagger}_{-k}\right)
+\frac{Z_{I}}{M_{I}}\left(\mathbf{p}_{I}\cdot\boldsymbol{\epsilon}^{*}_{\pm}\right)e^{-i\mathbf{k}\mathbf{R}_{I}}\left(b^{\dagger}_{k}+b_{-k}\right)\right]\\
+&\frac{\lambda^{2}}{2\omega_{k}}\left(N_{e}+\frac{Z^{2}_{I}}{2\omega_{k}}\right)\left(b_{k}+b^{\dagger}_{-k}\right)\left(b_{-k}+b^{\dagger}_{k}\right)+\omega_{k} (b^{\dagger}_{k}b_{k}+b^{\dagger}_{-k}b_{-k}+1),
\end{split}
\end{equation}
where we have used that $\epsilon_{0}=\frac{1}{4\pi}$ and atomic units, i.e. $Z_{i}=1$,  $\hbar=1$ and $m_{i}=1$, have been adopted.
The field squared term can be removed using a Bogoljubov transformation \cite{anappara2009signatures}
\begin{equation}
U_{k} = \textrm{exp}\left[\theta\left(b^{\dagger}_{k}b^{\dagger}_{-k}-b_{k}b_{-k}\right)\right],   
\label{eq:Rotation}
\end{equation}
where $\tanh{2\theta}=\frac{N_{e}\lambda^{2}}{N_{e}\lambda^{2}+2\omega^{2}}$. In that case, the Hamiltonian in Eq.\ref{eq:Multimode} becomes equal to 
\begin{equation}
\begin{split}
\textrm{H}_{\pm} &=\sum_{i}\frac{p^{2}_{i}}{2}+\sum_{I}\frac{p^{2}_{I}}{2M_{I}}-\sum_{i, I}\frac{Z_{I}}{\left|r_{i}-R_{I}\right|}+\sum_{i>j}\frac{1}{\left|r_{i}-r_{j}\right|}+\sum_{I>J}\frac{Z_{I}Z_{J}}{\left|R_{I}-R_{J}\right|}\\
-&\sum_{i}\frac{\lambda \left(\mathbf{p}_{i}\cdot\boldsymbol{\epsilon}_{\pm}\right)e^{i\mathbf{k}\mathbf{r}_{i}}}{\sqrt{2\omega_{k}}}\left(b_{k}+b^{\dagger}_{-k}\right)
-\sum_{i}\frac{\lambda \left(\mathbf{p}_{i}\cdot\boldsymbol{\epsilon}^{*}_{\mp}\right)e^{-i\mathbf{k}\mathbf{r}_{i}}}{\sqrt{2\omega_{k}}}\left(b^{\dagger}_{k}+b_{-k}\right)\label{eq:Multimode_1}
\\  
+&\sum_{I}\frac{Z_{I}\lambda \left(\mathbf{p}_{I}\cdot\boldsymbol{\epsilon}_{\pm}\right)e^{i\mathbf{k}\mathbf{R}_{I}}}{M_{I}\sqrt{2\omega_{k}}}\left(b_{k}+b^{\dagger}_{-k}\right)
+\sum_{I}\frac{Z_{I}\lambda \left(\mathbf{p}_{I}\cdot\boldsymbol{\epsilon}^{*}_{\mp}\right)e^{-i\mathbf{k}\mathbf{R}_{I}}}{M_{I}\sqrt{2\omega_{k}}}\left(b^{\dagger}_{k}+b_{-k}\right)\\
+&\frac{N_{e}\lambda^{2}}{2\omega_{k}}\left(b_{k}+b^{\dagger}_{-k}\right)\left(b_{-k}+b^{\dagger}_{k}\right)+\sum_{I}\frac{Z^{2}_{I}\lambda^{2}}{2\omega_{k}M_{I}}\left(b_{k}+b^{\dagger}_{-k}\right)\left(b_{-k}+b^{\dagger}_{k}\right)\\
+&\omega_{k} (b^{\dagger}_{k}b_{k}+b^{\dagger}_{-k}b_{-k}+1),   
\end{split}    
\end{equation}
with 
\begin{equation}
\tilde{\omega}=\sqrt{\omega^{2}+\lambda^{2}\left(N_{e}+\sum_{I}\frac{Z^{2}_{I}}{M_{I}}\right)}.    
\end{equation}
Introducing the new bosonic operators 
\begin{equation}
\alpha = \frac{b_{k}+b_{-k}}{\sqrt{2}} \hspace{1cm} \beta = \frac{b_{k}-b_{-k}}{\sqrt{2}}   
\label{eq:photons}
\end{equation}
the starting point for our theoretical investigations is obtained
\begin{equation}
\begin{split}
\textrm{H}_{\pm} =&\sum_{i}\frac{p^{2}_{i}}{2}+\sum_{I}\frac{p^{2}_{I}}{2M_{I}}+\tilde{\omega} (\alpha^{\dagger}\alpha+\beta^{\dagger}\beta+1) \\
-&\sum_{i}\Delta_{i}\left(\alpha+\alpha^{\dagger}\right)-\sum_{i}\chi_{i}\left(\beta^{\dagger}-\beta\right)\\  
+&\frac{Z_{I}}{M_{I}}\sum_{I}\Delta_{I}\left(\alpha+\alpha^{\dagger}\right)+\frac{Z_{I}}{M_{I}}\sum_{I}\chi_{I}\left(\beta^{\dagger}-\beta\right)\\
+&\sum_{i>j}\frac{1}{\left|r_{i}-r_{j}\right|}+\sum_{I>J}\frac{Z_{I}Z_{J}}{\left|R_{I}-R_{J}\right|}\label{eq:Multimode_2}\\
-&\sum_{i, I}\frac{Z_{I}}{\left|r_{i}-R_{I}\right|},
\end{split}
\end{equation}
where
\begin{align}
&\Delta=\frac{\lambda}{\sqrt{2\tilde{\omega}}}(p_{x}\cos(kr)-p_{y}\sin(kr))\label{eq:2}\\
&\chi =\frac{i\lambda}{\sqrt{2\tilde{\omega}}}(p_{x}\sin(kr)+p_{y}\cos(kr)).\label{eq:1}
\end{align}
We note that Eq.\ref{eq:Multimode_2} becomes real if we perform the unitary rotation $W$ of the Hamiltonian, where $W$ equals 
\begin{equation}
W = \textrm{exp}\left(i\frac{\pi}{2}\alpha^{\dagger}\alpha\right).    
\end{equation}
From Eq.\ref{eq:Multimode_2} we observe that strong coupling to a quantized field entangles electrons, photons and vibrations. However, in this work, we focus exclusively on resonances with electronic excitations. This justifies the adoption of the Born-Oppenheimer approximation where the nuclei are fixed in space and their motion is therefore neglected \cite{flick2017cavity,angelico2023coupled,fiechter2024understanding,schnappinger2023cavity}. We note that considering the fixed nuclei also means that their coupling to the field is neglected. The vibrational strong coupling regime as well as the interplay between electronic and vibrational components will be the focus of a following study. 

\subsection{Symmetries of a molecule in a chiral cavity}
\begin{figure}
    \centering
    \includegraphics[width=\textwidth]{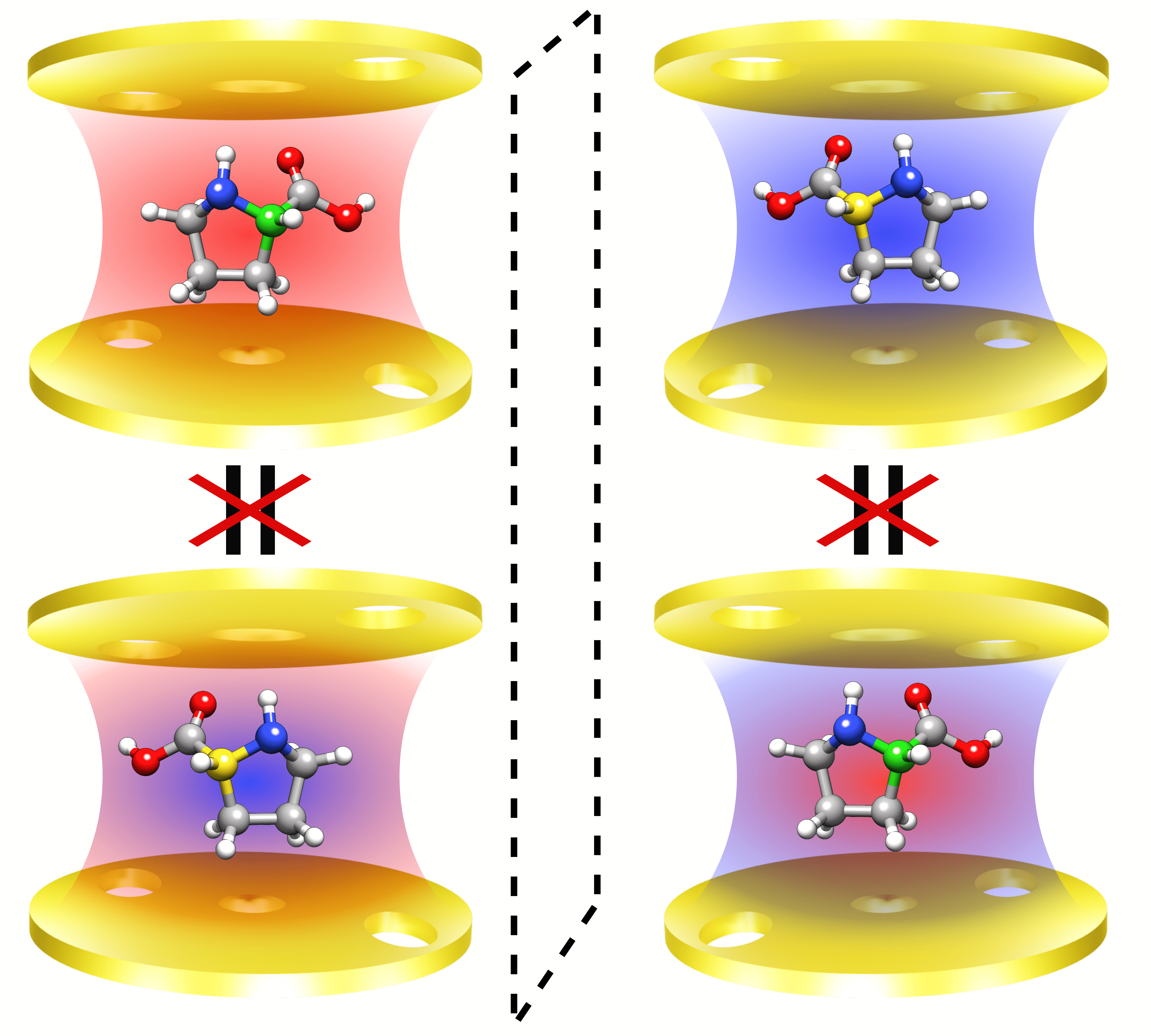}
    \caption{\textbf{Symmetries of a chiral molecule in a chiral cavity.} While the chiral cavity lifts the energy degeneracy between the two enantiomers of a chiral molecule (left and right column), inverting both the field polarization and the enantiomer in the cavity (full system reflection, upper and lower stripe) leads to degenerate systems.}
    \label{fig:symmetry}
\end{figure}

Outside the cavity, the two mirror images of a chiral molecule have the same energy levels. This is because the electronic Hamiltonian does not include any parity-violating terms. 
The chiral cavity introduces an asymmetry that breaks the equivalence between the enantiomers.
However, the Hamiltonian in Eq.\ref{eq:Multimode} does not include any parity-violating term for a global reflection of the whole system, cavity and molecule. A global reflection of the molecule and the cavity leads to a change in the enantiomer and in the circular polarization of the field, and therefore produces a system with the same energy levels as the initial one. This shows a critical and interesting symmetry of the Hamiltonian, pictorially illustrated in Fig.\ref{fig:symmetry}: inverting the chirality of the cavity and inverting the molecular chirality are equivalent operations. 
\section{Coupled cluster formalism in chiral cavities}
To describe the interaction between the field and the electrons we will use a coupled cluster based method\cite{haugland2020coupled,liebenthal2023assessing}. The electron-photon wave function is parametrized as
\begin{eqnarray}
\ket{\psi} =\textrm{exp}\left(T_{1}+T_{2}+(S_{\alpha}+\gamma_{\alpha})\alpha^{\dagger}+(S_{\beta}+\gamma_{\beta})\beta^{\dagger}\right)\ket{\textrm{HF}}\otimes\ket{0,0},  \label{eq:QED-CCSD-S-MINIMAL}
\end{eqnarray}
where $\ket{\textrm{HF}}$ is the vacuum Hartree-Fock Slater determinant and $\ket{0,0}$ denotes the photonic vacuum for the $\alpha$ and $\beta$ bosons introduced in Eq.~(\ref{eq:photons}). 
The excitation operators in the exponential of Eq.~(\ref{eq:QED-CCSD-S-MINIMAL}) allow us to include correlation. They can be divided into three different categories:  electron ($T_{1}$ and $T_{2}$), photon ($\gamma_{\alpha}$ and $\gamma_{\beta}$) and electron-photon ( $S^{\alpha}\alpha^{\dagger}$ and $S^{\beta}\beta^{\dagger}$) excitation operators. In second quantization, these operators are
\begin{align}
&T_{1} = \sum_{ai}t^{a}_{i}E_{ai} \nonumber\\
&T_{2}=\frac{1}{2}\sum_{abij} t^{ab}_{ij}E_{ai}E_{bj}\nonumber\\
&S_{\alpha}=\sum_{ai}s^{a}_{i\alpha}E_{ai}+ \frac{1}{2}\sum_{aibj}s^{ab}_{ij\alpha}E_{ai}E_{bj}\nonumber\\
&S_{\beta}=\sum_{ai}s^{a}_{i\beta}E_{ai}+ \frac{1}{2}\sum_{aibj}s^{ab}_{ij\beta}E_{ai}E_{bj},\label{eq:Excitation}
\end{align}
where the singlet operator $E_{pq}$ has been introduced
\begin{equation}
E_{pq} = \sum_{\sigma}a^{\dagger}_{p\sigma}a_{q\sigma}.\label{eq:Second_quantized}    
\end{equation}
In Eq. \ref{eq:Second_quantized}, $a^{\dagger}_{p\sigma}$ creates and $a_{p\sigma}$ annihilates an electron in orbital $p$ with spin $\sigma$. The indices $i,j$ and $a,b$ label occupied and empty orbitals in the HF reference, respectively.
The parameters $t^{a}_{i}$, $t^{ab}_{ij}$, $s^{a}_{i\alpha}$, $s^{a}_{i\beta}$ as well as $\gamma_{\alpha}$ and $\gamma_{\beta}$ are referred to as amplitudes and they are determined in the ground state calculations. When the full set of electronic and photonic excitations is included in the cluster operator 
\begin{eqnarray}
T=\sum_{p}T_{p}+\sum_{n}\sum_{m}\left(\gamma^{n\alpha,m\beta}+\sum_{p}S^{n\alpha,m\beta}_{p}\right)\alpha^{\dagger n}\beta^{\dagger m},
\end{eqnarray}
the coupled cluster method becomes exact within the basis set chosen for the electrons. 
Truncation of the excitation manifold is however needed in practical calculations and in this work we truncate the excitation manifold to include up to double excitations in the electrons and up to single excitations in the photonic space. The method is therefore named minimal coupling quantum electrodynamics coupled cluster singles and doubles (MC-QED-CCSD). 
The wave function optimization is achieved by requiring that the projections of the Schr\"{o}dinger equation on the excitations featured in the cluster operator are equal to zero
\begin{eqnarray}
\Omega_{\mu n m}=\bra{\mu,n,m}\bar{H}\ket{\textrm{HF},0,0}=0,
\label{eq:Omega}
\end{eqnarray}
where
\begin{align}
&\ket{\textrm{HF},0,0}  = \ket{\textrm{HF}}\otimes\ket{0,0}\nonumber\\[3pt]
&\ket{\mu,n,m}=\ket{\mu}\otimes\ket{m,n}  
\end{align}
and the cluster operator is
\begin{align}
 &T = T_{1}+T_{2}+(S_{\alpha}+\gamma_{\alpha})\alpha^{\dagger}+(S_{\beta}+\gamma_{\beta})\beta^{\dagger}.
\end{align}
The similarity transformed Hamiltonian has also been introduced
\begin{equation}
\bar{H}=\textrm{exp}\left(-T\right)H\;\textrm{exp}\left(T\right). 
\end{equation}
In Equation \ref{eq:Omega}$, \mu$ labels an electronic excitation while $n$ and $m$ are photonic excitations in $\alpha$ and $\beta$. The ground state energy, $E_g$, is equal to
\begin{align}
E_{g} =& \bra{\textrm{HF},0,0}\bar{H}\ket{\textrm{HF},0,0}\label{eq:energy}\\
\hspace{6mm}=& E_{\textrm{HF}} + \sum_{aibj}(t^{ab}_{ij}+t^{a}_{i}t^{b}_{j})(2g_{iajb}-g_{ibja})\nonumber\\
\hspace{5.5mm}-&\sum_{ai}\Delta_{ia}(s^{a}_{i\alpha} + \gamma_{\alpha} t^{a}_{i})-\sum_{ai}   \chi_{ia} (s^{a}_{i\beta} + \gamma_{\beta} t^{a}_{i}),\nonumber 
\end{align}
where $E_{\textrm{HF}}$ is the HF energy and the $g_{pqrs}$ are the two electron integrals \cite{helgaker2014molecular}. For further details on the ground state optimization, we refer the reader to Refs.\citenum{riso2023strong} and \citenum{riso2023strongg}.
\subsection{MC-QED-CCSD: Excited states}
To compute the excited states of a system inside the chiral cavity we extend the equation of motion formalism (EOM) to our MC-QED-CCSD\cite{haugland2020coupled,pavosevic2021polaritonic,liebenthal2022equation,riso2022characteristic}. Within this approach, the excitation energies are computed by diagonalizing the similarity transformed Hamiltonian $\bar{H}$ in the space $\{\ \ket{\textrm{HF},0,0},\ket{\mu,n,m}\}$
\begin{eqnarray*}
\cr\bar{H}&= \left( \begin{array}{cc}
 \bra{HF,0,0}\bar{H}\ket{HF,0,0} & \bra{HF,0,0}\bar{H}\ket{\mu,m,n}\\
    \bra{\mu,m,n}\bar{H}\ket{HF,0,0} & \bra{\nu,p,q}\bar{H}\ket{\mu,m,n}
\end{array} \right)  \\
&= \left(\begin{array}{cc}E_{g} & \eta_{\mu n m} \nonumber\\
    \Omega_{\mu  n  m} & A^{\nu p q}_{\mu n m}+ E_{g}\delta_{\mu \nu}\delta_{m p}\delta_{n q}\nonumber\end{array}\right),
\end{eqnarray*}
where the Jacobian matrix $\mathbf{A}$ is equal to \cite{haugland2020coupled}
\begin{equation}
 A^{\nu p q}_{\mu n m} = \bra{\nu p q}\left[\bar{H},\tau^{\dagger}_{\mu n m}\right] \ket{\textrm{HF},0,0}.   
\end{equation}
The operator $\tau^{\dagger}_{\mu n m}$ creates $\ket{\mu,n,m}$ if applied to $\ket{\textrm{HF},0,0}$. Since coupled cluster methodologies are non-Hermitian, the left and right eigenvectors of $\mathbf{A}$ are not identical. We therefore have two sets of eigenvectors: right eigenvectors $\ket{R_{n}}$ and left eigenvectors $\bra{L_{n}}$ defined as 
\begin{align}
\ket{R_{n}}=& \left(R^{n}_{0}\ket{HF,0,0}+\sum_{\mu}R^{n}_{\mu}\ket{\mu,0,0}+\sum_{\mu}R^{n\alpha}_{\mu}\ket{\mu,1,0}\right.\nonumber\\
+&\left.\sum_{\mu}R^{n\beta}_{\mu}\ket{\mu,0,1}+R^{n\alpha}_{0}\ket{HF,1,0}+R^{n\beta}_{0}\ket{HF,0,1}\right),  \\
\bra{L_{n}}=& \left(\bra{HF,0,0}L^{n}_{0}+\sum_{\mu}\bra{\mu,0,0}L^{n}_{\mu}+\sum_{\mu}\bra{\mu,1,0}L^{n\alpha}_{\mu}\right.\nonumber\\
+&\left.\sum_{\mu}\bra{\mu,0,1}L^{n\beta}_{\mu}+\bra{HF,1,0} L^{n\alpha}_{0}+\bra{HF,0,1} L^{n\beta}_{0}\right),    
\end{align}
where $R^{n}_{\mu}$ and $L^{n\alpha}_{\mu}$ are state specific coefficients. 
The EOM scheme provides accurate size-intensive excitation energies. The excited state transition properties are obtained as the square root of the properties of the left and right eigenvectors. For example, the transition density of the $n^{th}$ excited state is equal to
\begin{eqnarray}
\rho_{n}(r)=\sum_{pq}\psi^{*}_{p}(r)\psi_{q}(r)\sqrt{\bra{\Lambda}E_{pq}\ket{R_{n}}\bra{L_{n}}E_{pq}\ket{\psi}},    
\end{eqnarray}
where $\bra{\Lambda}$ is the left side equivalent of the coupled cluster ground state and $\psi^{*}_{p}(r)$ and $\psi_{q}(r)$ are the HF molecular orbitals\cite{helgaker2014molecular}. 
\subsection{Jaynes Cummings model for chiral cavities}
In the Jaynes Cummings (JC) model, molecules are approximated as two-level systems with a ground state $\ket{G}$ and an excited state $\ket{E}$\cite{larson2021jaynes}. 
The energies of these states are denoted as $E_{g}$ and $E_{e}$ respectively.
The interaction of a two-level system with the field is accounted for in the rotating wave approximation (RWA), meaning that only energy-conserving elements of $\mathbf{p}\cdot\mathbf{A}$ are retained, i.e. photon creation-molecular de-excitation or photon annihilation-molecular excitation. The Hamiltonian for a single molecule in strong coupling with the chiral field therefore becomes 
\begin{align}
H =& \omega (\alpha^{\dagger}\alpha+\beta^{\dagger}\beta) + E_{g}\ket{G}\bra{G} + E_{e}\ket{E}\bra{E} \nonumber\\
-&\Delta_{eg}\alpha\ket{E}\bra{G}-\Delta_{ge}\alpha^{\dagger}\ket{G}\bra{E}\label{eq:JC}\\
-&\chi_{eg}\beta\ket{E}\bra{G}-\chi_{ge}\beta^{\dagger}\ket{G}\bra{E},\nonumber   
\end{align}
where $\Delta_{eg}$ and $\chi_{eg}$ are the transition moments between the excited state and the ground state, see Eq.\ref{eq:2} and Eq.\ref{eq:1}.  
We note that using the JC approach, the field-induced modifications on the electronic ground state are completely neglected. The electronic parameters to be inserted in the JC model are usually obtained through an electronic structure calculation without the cavity. We compute these quantities at the CCSD level. Because of the non-Hermiticity of coupled cluster approach, the JC Hamiltonian needs to be rewritten as
\begin{align}
H =& \omega (\alpha^{\dagger}\alpha+\beta^{\dagger}\beta) + E_{g}\ket{G}\bra{G} + E_{e}\ket{E}\bra{E} \nonumber\\
-&\sqrt{\Delta_{L_{1}CC}\Delta_{\Lambda R_{1}}}\alpha\ket{E}\bra{G}-\sqrt{\Delta_{L_{1}CC}\Delta_{\Lambda R_{1}}}\alpha^{\dagger}\ket{G}\bra{E}\label{eq:JC_CC}\\
-&\sqrt{\chi_{L_{1}CC}\chi_{\Lambda R_{1}}}\beta\ket{E}\bra{G}-\sqrt{\chi_{L_{1}CC}\chi_{\Lambda R_{1}}}\beta^{\dagger}\ket{G}\bra{E}.\nonumber   
\end{align}
Extension of the JC approach to n molecules is obtained by using multimolecular excited states 
\begin{align}
&\ket{G^{1},G^{2},G^{3},G^{4},...,G^{n}}\equiv\ket{G};  \\
&\ket{R^{1}_{1},G^{2},G^{3},G^{4},...,G^{n}}\equiv\ket{R^{1}_{1}};  \\
&\ket{G^{1},R^{2}_{1},G^{3},G^{4},...,G^{n}}\equiv\ket{R^{2}_{1}}\;\;\;etc,
\end{align}
following the Tavis-Cummings procedure \cite{davidsson2023role,larson2021jaynes}.

\section{Results}
In this section, we use the methodologies presented in the theory section to understand the effects that strong coupling to circularly polarized fields has on the excited states of a chiral molecule. The \textit{ab initio} calculations have been performed using a development
version of the eT program \cite{folkestad20201}, and the molecular structures have been optimized using the ORCA software package \cite{neese2020orca} at the DFT-B3LYP/def2-SVP level \cite{weigend2005a,schuchardt2007a,feller1996a,pritchard2019a}. The Jaynes-Cummings calculations have been performed using python scripts that can be found in the Zenodo link \citenum{Zenodo}.

\subsection{Single molecule calculations}
\begin{figure}
    \centering
    \includegraphics[width=\textwidth]{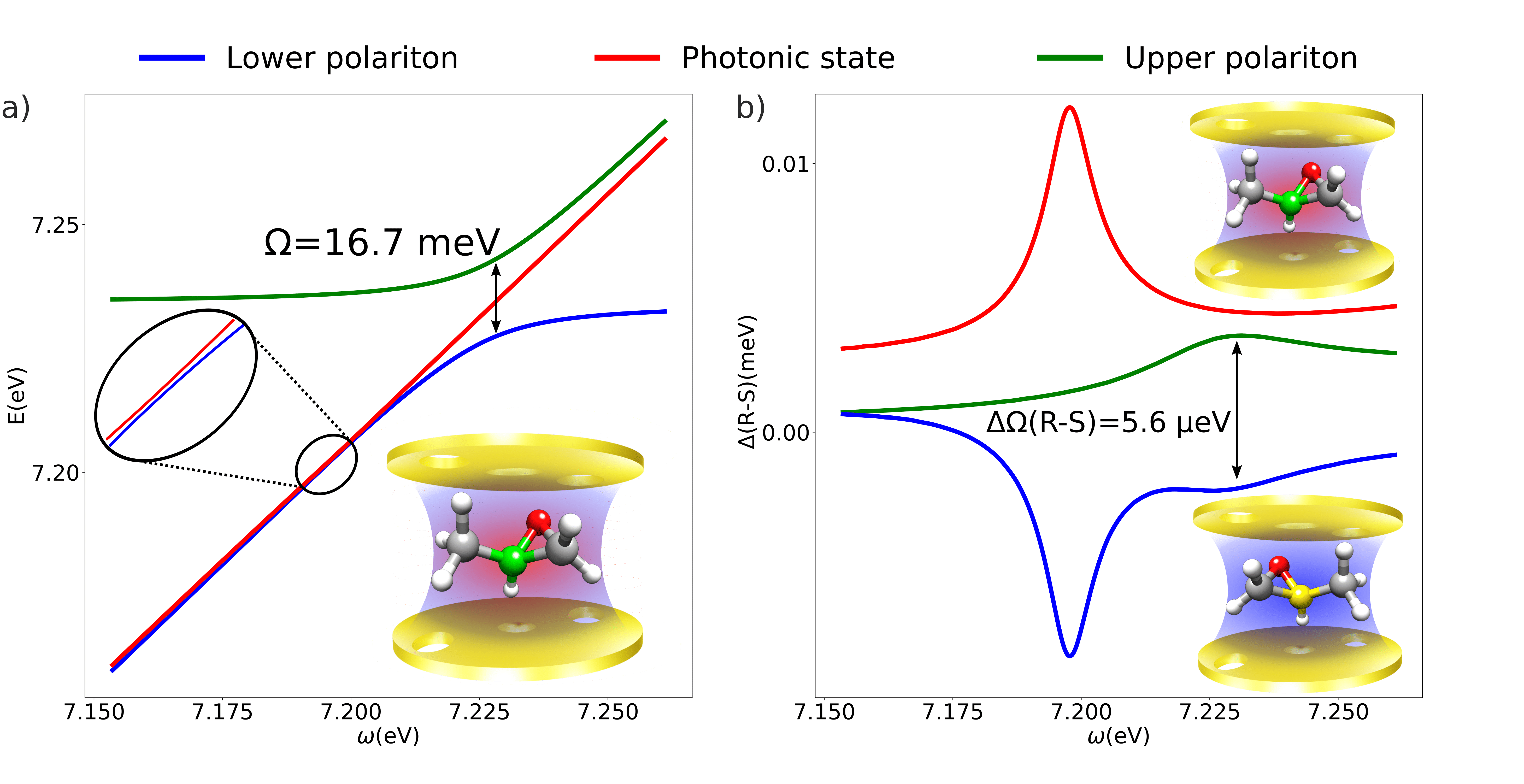}
    \caption{a)Dispersion of the excitation energies for S-methyloxirane in a RHCP chiral cavity. An avoided crossing is observed when the electronic and photonic excitation energies coincide and also when the photonic energies are nearly degenerate, see inset. b) Field-induced discrimination in the excitation energies of the system.}
    \label{fig:One_mol}
\end{figure}
Methyloxirane is a small chiral compound very well studied in the literature for its chiro-optical properties\cite{fuse2022anharmonic}. Strong light-matter coupling can induce significant modifications of the molecular excitation energies. This phenomenon is most prominent when the cavity is resonant or nearly resonant with a molecular excitation. In this case, the coherent superposition of the two degenerate states leads to the formation of upper and lower polaritons (UP and LP respectively), and an avoided crossing is observed, see Fig.\ref{fig:One_mol}a. The energy difference between the polaritons at exact resonance is known as Rabi splitting $\Omega$. In a simplified Jaynes-Cummings picture, see Eq.\ref{eq:JC_CC}, the Rabi splitting amounts to $2\sqrt{\Delta_{eg}\Delta_{ge}+\chi_{eg}\chi_{ge}}$. The results in Fig.\ref{fig:One_mol} are however computed with MC-QED-CCSD.
Since chiral molecules preferentially absorb and emit one circular polarization of the field, i.e. the transition moments are different depending on the chosen enantiomer, enantiospecific Rabi splittings arise \cite{mauro2022charge,baranov2022towards}. In Fig.\ref{fig:One_mol}a, we plot the frequency dispersion of the excitation energies for a S-methyloxirane %\revER{(YOU MENTION HERE THE CHIRAL SYSTEM FOR THE FIRST TIME. I WOULD BENTION IT EARLIER WITH ALSO A SHORT CHEMICAL MOTIVATION OF WHY YOU CHOSE THIS SYSTEM)} 
in a RHCP cavity as computed using the  MC-QED-CCSD method. The basis set for the calculation is aug-cc-pVDZ \cite{dunning1989a,kendall1992a,woon1993a} and the coupling is set to $\lambda=0.005$ a.u. (quantization volume of around 100 $nm^{3}$). 
Since the cavity hosts two photonic modes, Fig.\ref{fig:One_mol} features three nearly degenerate excited states. 
In the decoupled setting, $\lambda=0$ a.u. , two of these states are photonic and one is instead the $S_0 \longrightarrow S_1$ methyloxirane transition.
At resonance, only one of the two cavity modes couples significantly with the molecular excitation, leading to the formation of a polaritonic pair (blue and green in Fig.\ref{fig:One_mol}).
Instead, the red state is mostly decoupled from the molecule. The Rabi splitting changes depending on the system's chirality from 16.770 meV for R-methyloxirane to 16.764 meV for S-methyloxirane, a change of 5.6$\;\mu$eV. 
\begin{figure}
    \centering
    \includegraphics[width=\textwidth]{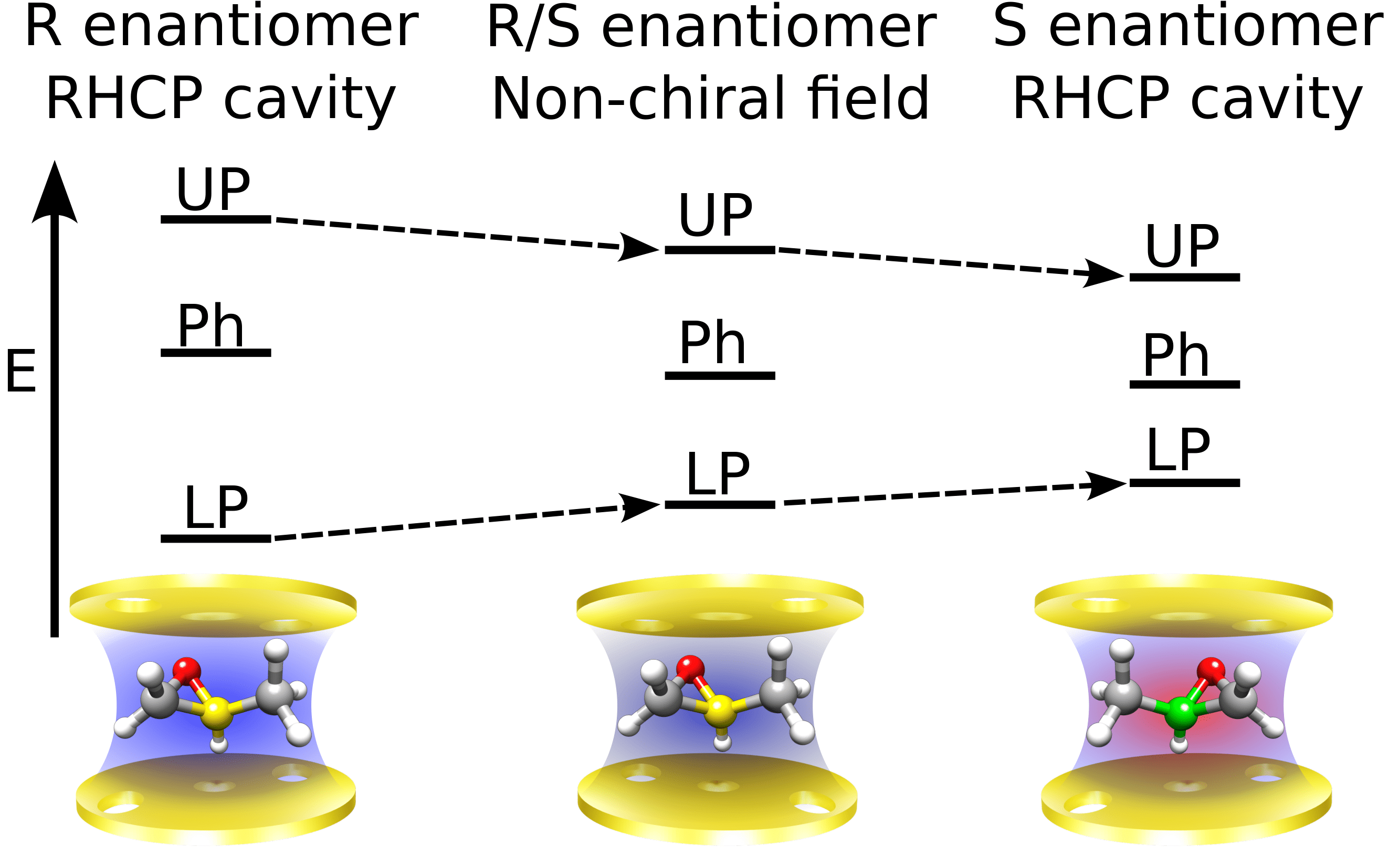}
    \caption{\textbf{Effect of the chiral field on the Rabi splitting of a chiral molecule.} Since methyloxirane is optically active, in a RHCP cavity the Rabi splitting for R and S-methyloxirane is not the same. If the field is non chiral (visually represented by the grey colour), instead, no differentiation between the enantiomers is observed.}
    \label{fig:Simple_effect}
\end{figure}
The difference in the dispersion for the R and S enantiomers, $\Delta(R-S)$, is displayed in Fig.\ref{fig:One_mol}b. We notice that the field-induced discrimination are in the $\mu$eV energy range and that for each state (LP, UP or photonic) the sign of the chiral effects remains the same for a wide range of frequencies. Specifically, in the configuration of Fig.\ref{fig:One_mol}, the lower polariton excitation energy is smaller for the R enantiomer while the opposite behavior is observed for the upper polariton and the photonic state. Figure \ref{fig:Simple_effect} showcases a simplified representation of the chiral field effects.  
%\revMC{I am a bit confused by this sentence, I think it's the definition of 'states that stabilize' rather than 'states that are stabilized'. Before the comma is the state that is stabilized, and after is the state that stabilizes S or R. Maybe it's better to compare the energies with respect to the non-chiral field case?}. 
Interestingly, the maximum field-induced discrimination is reached before resonance between the molecular excitation and cavity field. More precisely, the maximum in the field-induced discrimination is observed at an avoided crossing between the almost photonic states at $\omega\approx$ 7.2 eV, see inset of Fig.\ref{fig:One_mol}a. The two photonic states have different energies because the molecule interacts differently with the two field modes. 
Even more interesting is noticing that changing the enantiomer in the cavity modifies not only the energies of the photonic states but also their differences $\Delta E_{ph} $, see Fig.\ref{fig:Diff}. We stress that these are purely photonic states in an out-of-resonance setting. 
%\revER{In Fig.\ref{fig:Diff}, the maximum in energy difference is observed at the avoided crossing between the photonic states. (TO BE CLARIFIED AS IN COMMENT IN THE FIGURE CAPTION)}
\begin{figure}
    \centering
    \includegraphics[width=0.65\textwidth]{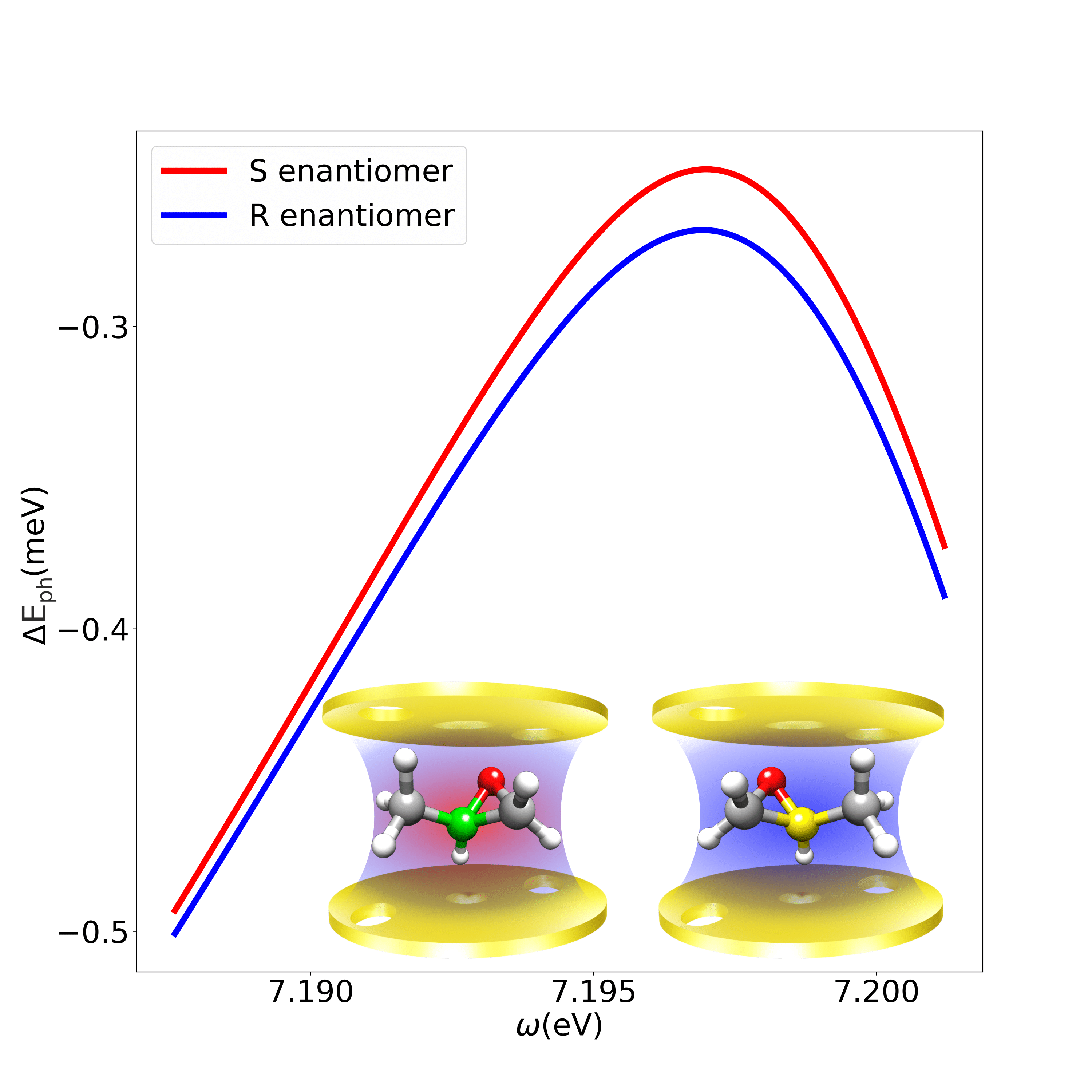}
    \caption{Difference in the energy of the purely photonic states in the chiral cavity of Fig.\ref{fig:One_mol}. The curve changes depending on the strongly coupled enantiomer because the photonic dispersion is modified by the interaction with the molecule. }%\revER{(IT IS NOT VERY CLEAR OR AT LEAST IMMEDIATE WHAT WE ARE LOOKING AT IN THIS FIGURE COMPARED TO THE PREVIOUS FIGURES!!)}}
    \label{fig:Diff}
\end{figure}  
Agreement between \textit{ab initio} and Jaynes Cummings calculations are quite good, with JC predicting a Rabi splitting of 14.9857 meV for the R enantiomer and 14.9800 meV for the S enantiomer. The predicted field-induced discrimination is therefore equal to $5.7\mu$eV. The difference in the predicted quantities is due to the fact that the JC scheme neglects non-resonant effects like the difference in energy of the two photonic lines. If the modification of the ground state wave function due to the cavity is included in the JC scheme, the JC results and the EOM-MC-QED-CCSD results are instead identical. 

\begin{figure}
    \centering
    \includegraphics[width=\textwidth]{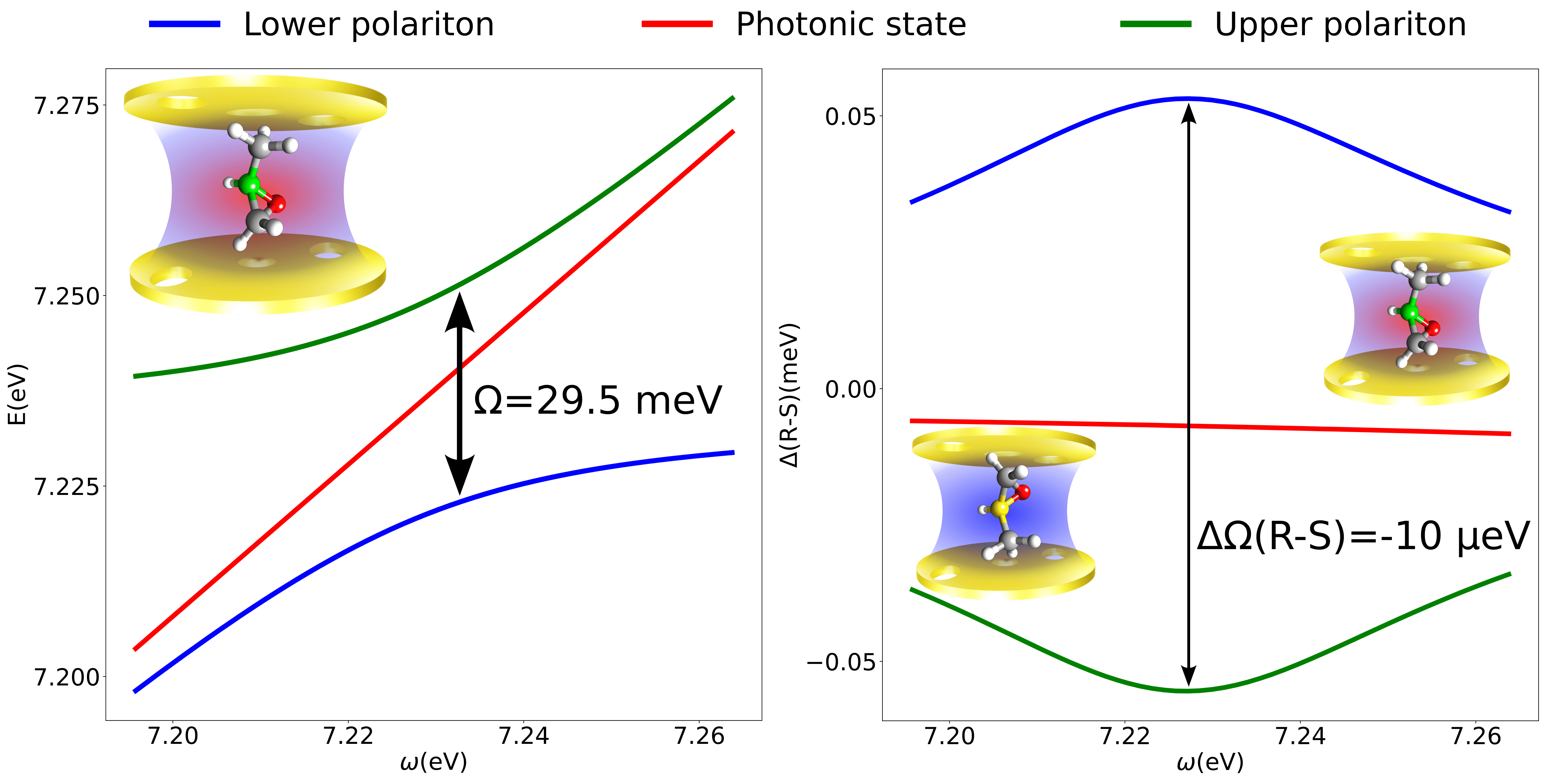}
    \caption{\textbf{Field induced effect at a different molecular orientation}. Compared to Fig.\ref{fig:One_mol}, the field discrimination exhibits a prominent sign change.}
    \label{fig:New_orientation}
\end{figure}

\begin{figure}
    \centering
    \includegraphics[width=\textwidth]{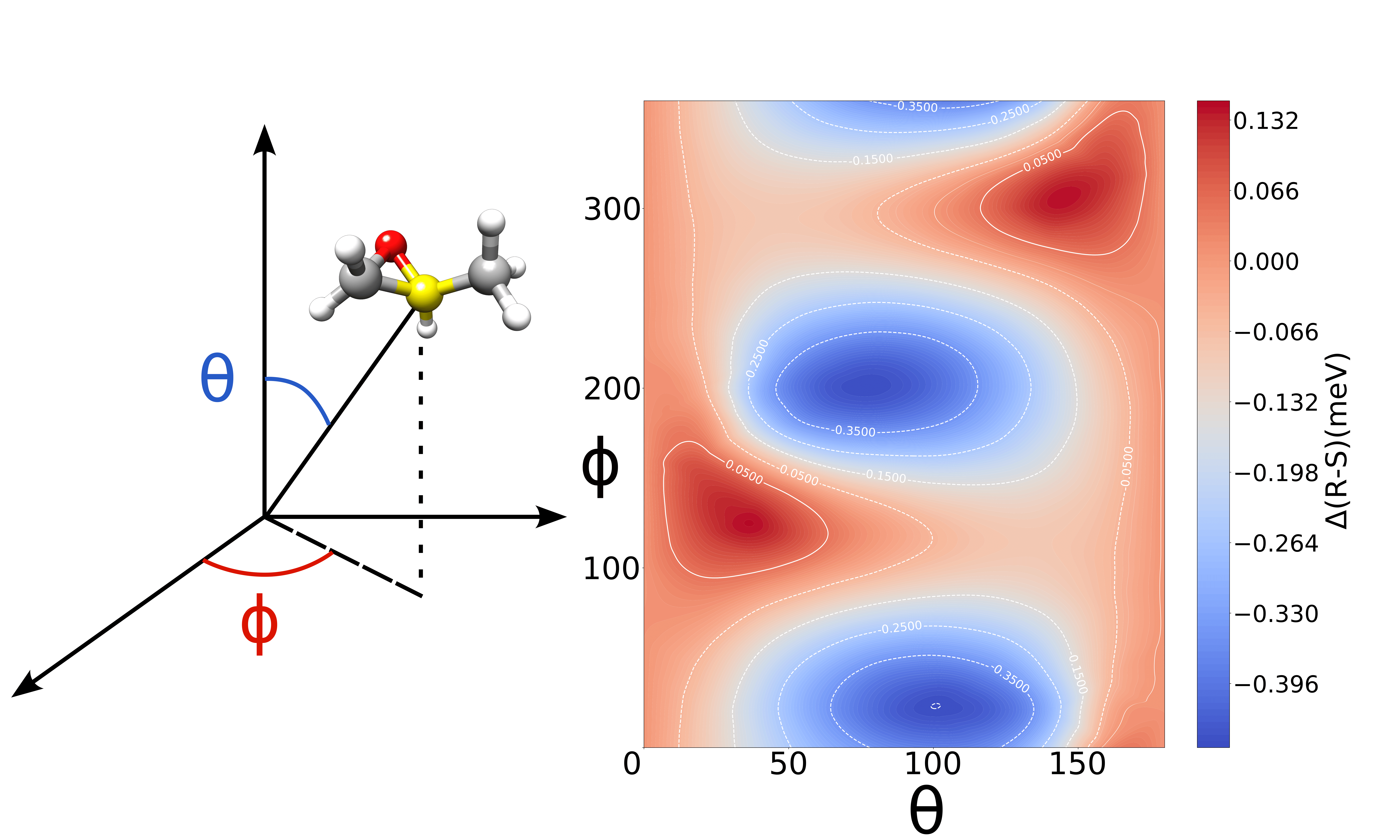}
    \caption{\textbf{Orientational effects on the Rabi splitting difference}. We notice that even after orientational averaging the field effects do not disappear.}
    \label{fig:Orientational}
\end{figure}

The cavity discrimination effect is heavily dependent on the molecular orientation. For example, in Fig.\ref{fig:New_orientation}, we show how the Rabi splitting and the field-induced discrimination change when the methyloxirane is rotated by 90$^{\circ}$ on the $y$ axis. A prominent change in the sign of the field-induced discrimination power is observed. Specifically, in this configuration the lower polariton stabilizes the S enantiomer while the upper polariton stabilizes the R enantiomer. \\
In a cavity the molecules are free to rotate, therefore orientational effects need to be taken into account. In Fig.\ref{fig:Orientational} we plot the difference in Rabi splitting for R- and S-methyloxiranes in a RHCP cavity, $\Delta(R-S)$. The molecular configurations are parametrized by the angles $\theta$ and $\phi$. 
For different molecular orientations, $\Delta(R-S)$ can have different signs. However, we notice that the two minima (blue) are roughly 3 times as deep as the two maxima (red). If the orientational effects are averaged, the observed Rabi splitting changes by -0.11 meV depending on the enantiomer in the cavity. Since methyloxirane exhibits circular dichroism, the discrimination is robust. 
\section{Collective effects}
\begin{figure}
    \centering
    \includegraphics[width=\textwidth]{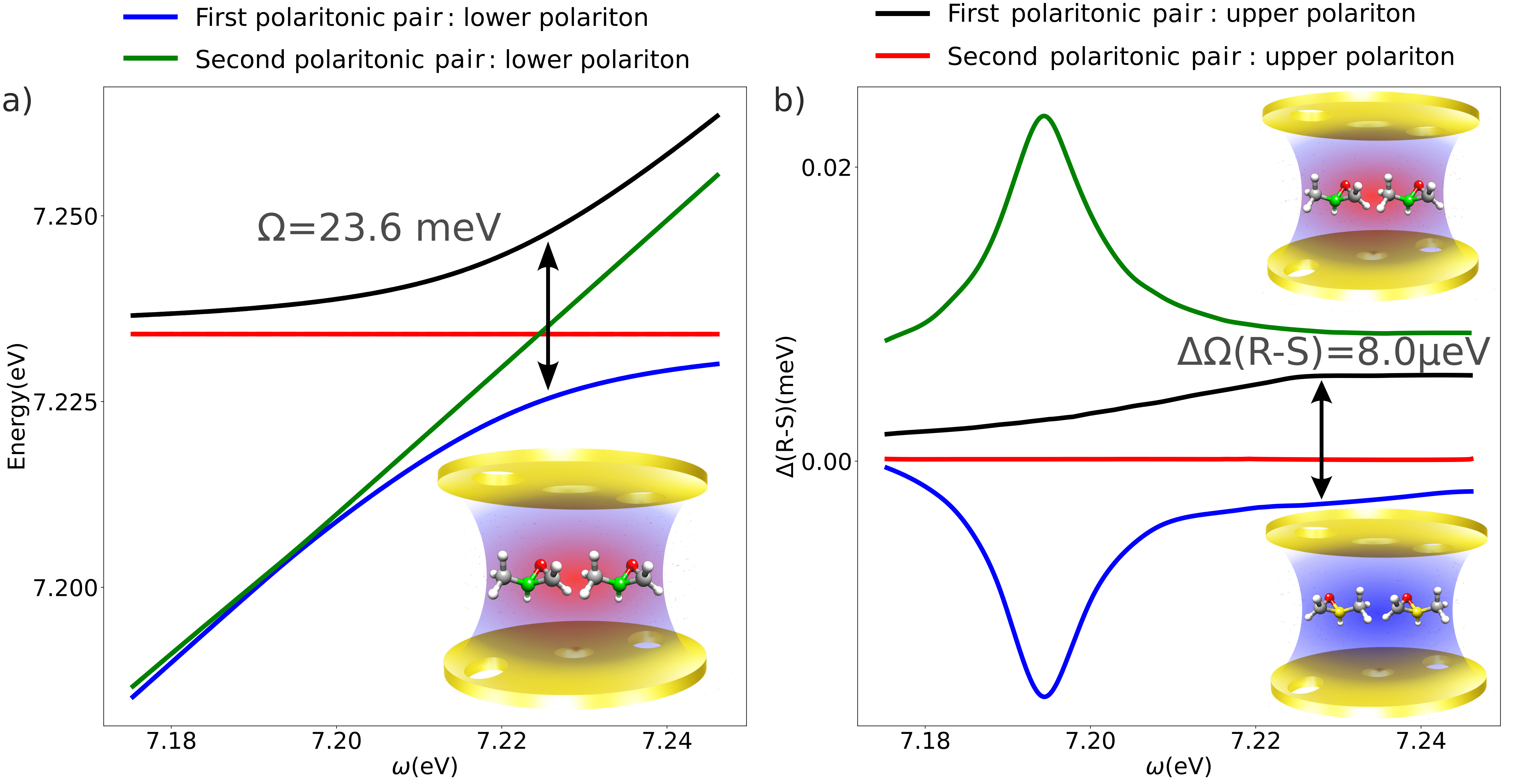}
    \caption{\textbf{Collective effects in the chiral cavity}. a) Dispersion of the excitation energies for two equal enantiomers displaced by 100\AA\; in a direction perpendicular to $\mathbf{k}$. b) Field-induced discriminating effects for the two-molecule case.}
    \label{fig:Two_mol}
\end{figure}

\begin{figure}
    \centering
    \includegraphics[width=\textwidth]{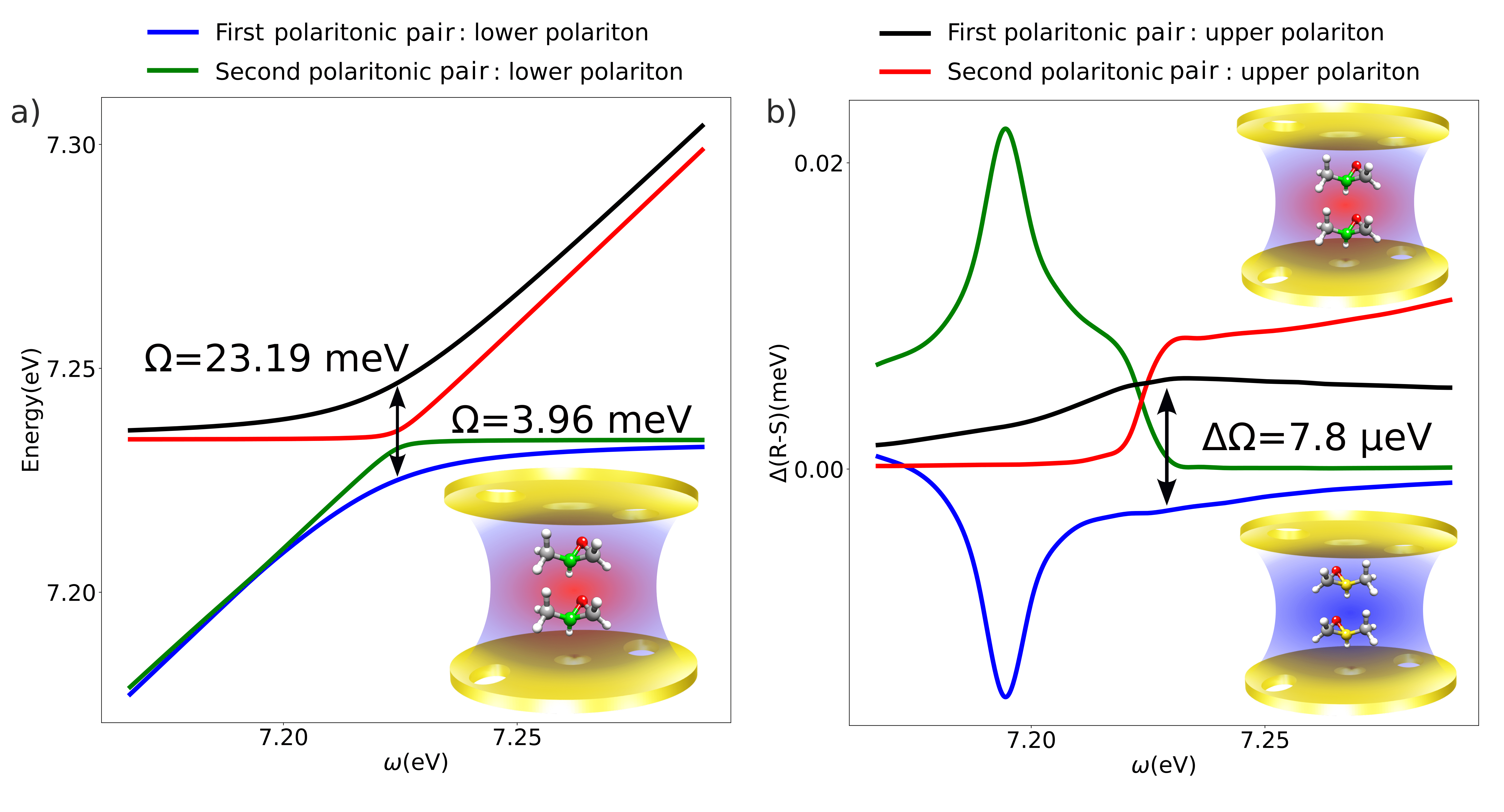}
    \caption{\textbf{Collective effects in the chiral cavity}. a) Dispersion of the excitation energies for two equal enantiomers displaced by 100 \AA\; along $\mathbf{k}$. b) Field-induced discriminating effects for the two-molecule case.}
    \label{fig:Two_mol_k}
\end{figure}

One of the most important features of the strong light-matter coupling phenomena is their collective behavior. As the number of molecules in the cavity grows, the Rabi splitting increases like the square root of the number of coupled systems. This collective behavior plays a critical role in making polaritonic effects observable for macroscopic systems. To check whether chiral discrimination properties also display a collective trend, we investigate the dispersion of the excitation energies for two identical enantiomers strongly coupled to the same field. To rule out electron-electron interactions, the two molecules are placed at a distance of 100 \AA\;. In Fig.\ref{fig:Two_mol}, the two molecules are shifted in the direction perpendicular to the field wave vector $\mathbf{k}$. We immediately notice that, in this case, two states do not significantly interact with the others: one electronic state (red) and one photonic state (green). In the following, we refer to the blue and black states as the first polaritonic pair and to the green and red states as the second polaritonic pair. 
We notice that the Rabi splitting in this new setting is $\Omega=23.6$ meV, which is $\sqrt{2}$ times larger than the Rabi splitting predicted for a single molecule in the cavity. The same $\sqrt{2}$ increase is also observed for the discrimination power of the field. When the two molecules are displaced in the direction perpendicular to the wave vector, the field-induced effects increase by $\sqrt{N}$ as is typical for strongly coupled systems. This is because all the molecules interact with the same field, i.e. the field phase is not changed. A detailed account of the deviation from the $\sqrt{N}$ trend when an extensive number of coupled molecules is considered has already been reported in Ref.\citenum{schafer2023chiral}. A different behavior should, however, be expected when the molecules are displaced along the $\mathbf{k}$ axis, see Fig.\ref{fig:Two_mol_k}.  
In Fig.\ref{fig:Two_mol_k}, we observe two polaritonic pair, one featuring the red and green states and one featuring the blue and black states.  A displacement in the $\mathbf{k}$ direction couples the red and green states and reduces slightly the splitting between the blue and black states. In this case, therefore, the Rabi splitting does not increase strictly following a collective effect but new avoided crossings are obtained. When looking at the field-induced discrimination, we notice that while the black and blue states maintain a constant stabilization sign, the states in the second polariton pair exhibit a change in the stabilization at their avoided crossing. This sign change can be attributed to a change in the nature of the state after the crossing. \\
\begin{figure}
    \centering
    \includegraphics[width=\textwidth]{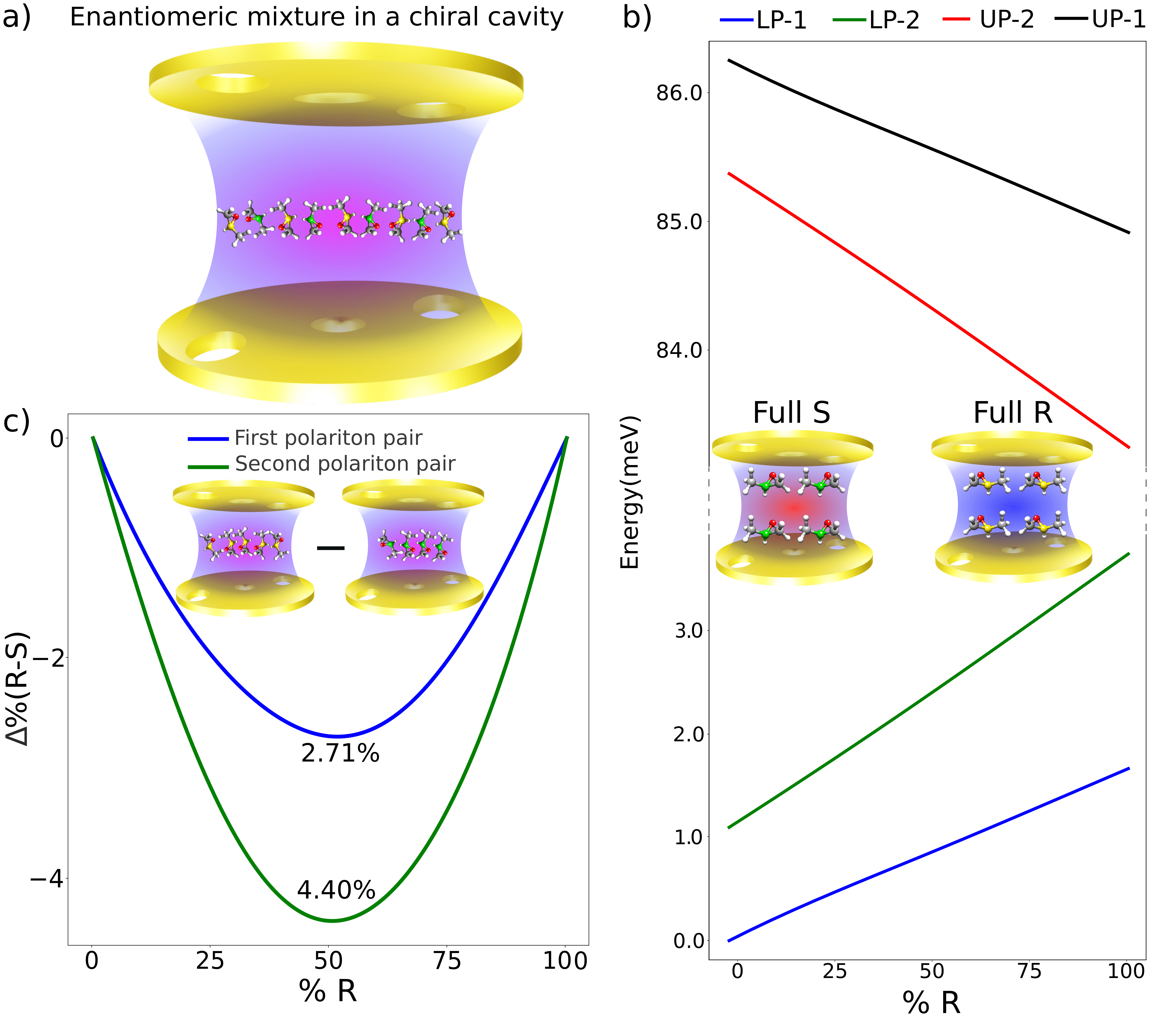}
    \caption{a) Collective behavior of flat methyloxirane molecules in a chiral cavity. b) Excitation energies %\revMC{The excitations are meV? or is this the rabi splitting?} 
    of the system as a function of the enantiomeric purity. The plot has been shifted such that the excitation energy of the lower polariton at a composition of 0$\% R$, is equal to zero. c) Population difference between R and S enantiomers in the polaritonic states and in the enantiomeric mixture.}
    \label{fig:methyloxirane}
\end{figure}
As discussed previously in the literature \cite{castagnola2024collective,sidler2020polaritonic,cui2022collective}, the field-induced effects observed in the experiments become apparent in the case of a large number of strongly coupled molecules. Furthermore, studying more than a few chiral systems at the same time allows us to account for the role that the enantiomeric purity of the strongly coupled sample has on the field-induced effects. The steep computational cost of the MC-QED-CCSD method, however, makes it unpractical for a sizeable number of molecules. Therefore, we use a Tavis-Cummings (TC) approach to discuss the effects of the chiral field on a multicomposite system. All the transition elements and energies used in the model are computed at the CCSD level as described in Eq.\ref{eq:JC_CC}. The system under study in Fig.\ref{fig:methyloxirane} is composed of a collection of 3720 methyloxiranes in their equilibrium geometries. The molecules can be either R or S enantiomers and they all lie in the same plane, i.e. they all interact with the field in phase, see Fig.\ref{fig:methyloxirane}a. Moreover, to eliminate orientational effects, each methyloxirane has a different orientation in space, (different $\theta$ and $\phi$ in Fig.\ref{fig:Orientational}.) %Studying a large number of chiral molecules allows us to account for the role that the enantiomeric purity of the sample has on the field-induced discrimination effects. 
%In \revER{Fig.\ref{fig:methyloxirane}, we display the chiral field effects on a collection of 3720 methyloxiranes in their optimized geometries in random orientations. All the molecules lie on the same plane, i.e. they all interact with the field in phase.
%\revER{
The basis set is aug-cc-pVDZ and the coupling strength is $\lambda=0.0005$ a.u. . 
In Fig.\ref{fig:methyloxirane}b, we see that changing the enantiomeric composition in the cavity has significant effects on the observed Rabi splitting. This statement is true for both polaritonic pairs. 
When the sample in the cavity is purely composed of the S enantiomer (0\% R), the Rabi splitting increases by 2 meV for the first polaritonic pair and 4 meV for the second polaritonic pair, as compared to the 100\% R case.
When the composition inside the cavity is not pure, instead, an intermediate splitting is found. The observed Rabi splittings can therefore be used to estimate the composition of the mixture. Because of circular dichroism, one of the two enantiomers preferentially interacts with the circularly polarized field. 
In this case, the stronger coupling is observed for the S enantiomer (the transition elements are, on average, larger). 
Therefore, we observe an asymmetry when we look at the percentage of the two enantiomers in the upper and lower polaritons. Given that the upper and lower polariton wave function is written as
\begin{align}
\ket{\psi}=&\ket{\underbrace{...G,...}_\text{R enantiomer},\underbrace{...G,...}_\text{S enantiomer},1_{ph}}C_{ph}+\sum_{i\in R \atop enantiomer}  \ket{\underbrace{...E,...}_\text{R enantiomer},\underbrace{...G,...}_\text{S enantiomer},0_{ph}}C_{i}\nonumber\\
+&\sum_{i\in S \atop enantiomer}  \ket{\underbrace{...G,...}_\text{R enantiomer},\underbrace{...E,...}_\text{S enantiomer},0_{ph}}C_{i}, 
\end{align}
the enantiomeric percentage in a polaritonic state is computed as
\begin{equation}
\% R/S = \sum_{i\in R/S \atop enantiomer}C^{2}_{i}.    
\end{equation}
Figure \ref{fig:methyloxirane}c shows how much the population of the polaritonic states deviates from the stoichiometry of the sample in the cavity. When the system is enantiomerically pure, the deviation is equal to zero while the population asymmetry reaches its maximum in the racemic mixture. This is an interesting outcome since the 50$\%$-50$\%$ mixture is, overall, achiral. Our results illustrate a basic example of field-induced excitation condensation on one type of enantiomers. Moreover, since the polaritons are richer in S enantiomer character, the dark states are richer in R-methyloxirane (precisely by 14 $\%$). These observations clearly show that the system's energy is not the only relevant parameter when dealing with symmetry breaking in the cavity. While in Fig.\ref{fig:methyloxirane} all the methyloxirane molecules are displaced on the axis perpendicular to the wave vector, the thickness in the $\mathbf{k}$ direction must also be accounted for in a realistic sample. In Fig.S1 (see Supplementary Material), we show how the discrimination effects change in the case of a droplet 15 nm long. No significant change in the field discrimination power is observed by changing the sample's $\mathbf{k}$ dimension.\\

\noindent The asymmetry effects discussed in Fig.\ref{fig:methyloxirane} can also be generated applying classical circularly polarized light on the molecular sample. In this case, the enantiomers will absorb circularly polarized light at different rates and different excited state populations will be observed. However, the cavity-induced discrimination is a vacuum effect, valid also when there are no real photons in the cavity. This is particularly relevant when dealing with excitations that can also be accessed thermally\cite{wang2019cavity}. An asymmetry in the enantiomeric dynamics is observed even if classical linearly polarized light is used to excite the sample in a chiral cavity. While the R and S enantiomers would absorb light at the same rate, for the methyloxirane case the R enantiomer will prominently follow the dynamics of the dark state while the S enantiomer will display more pronounced polaritonic modifications.\\

\section{Asymmetric energy redistribution in a chiral cavity}
\begin{figure}
    \centering
    \includegraphics[width=0.65\textwidth]{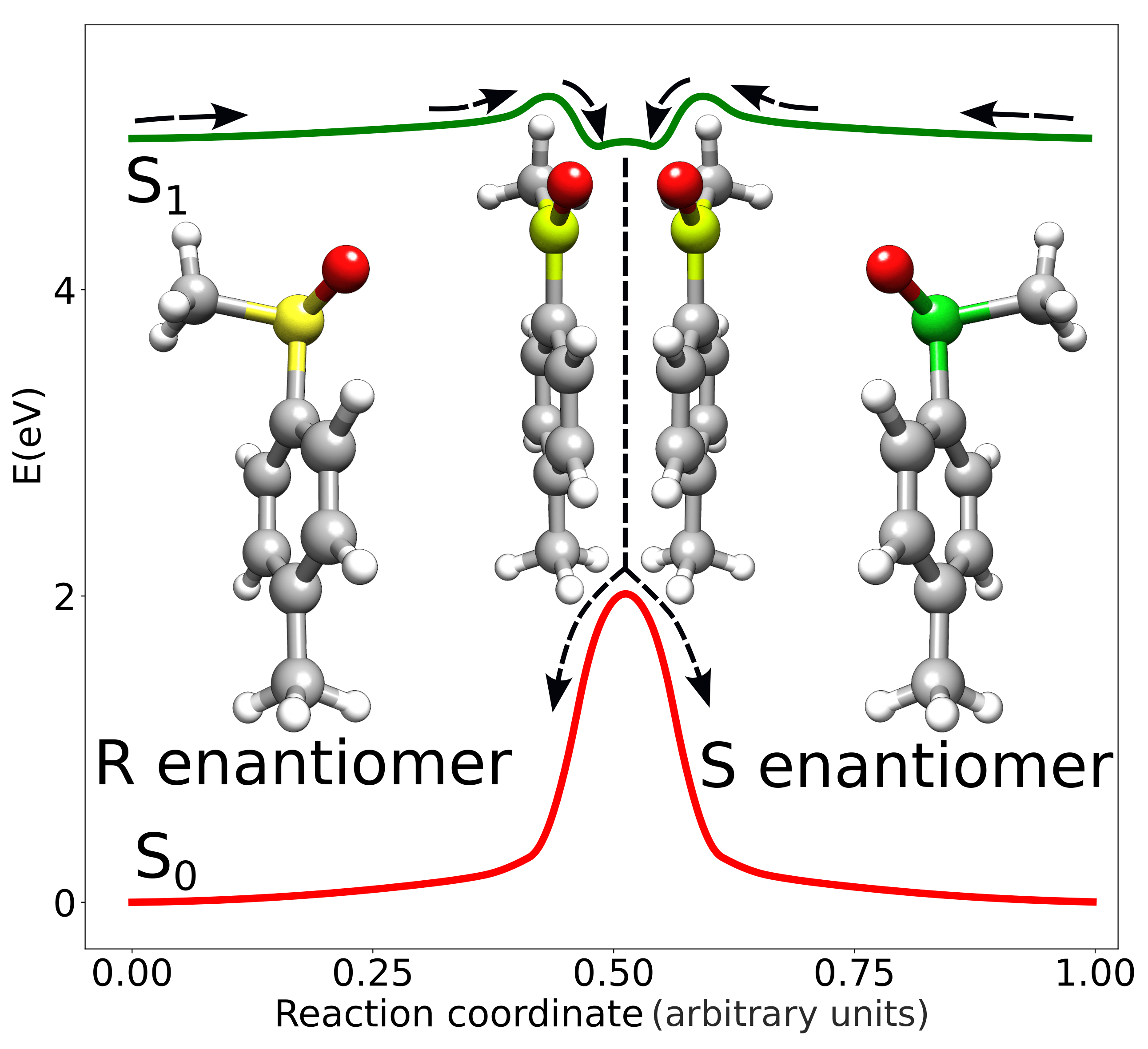}
    \caption{$S_{0}$ and $S_{1}$ potential energy surfaces for p-tolyl sulfoxide along the interconversion reaction coordinate.}
    \label{fig:enter-label}
\end{figure}
\begin{figure}
    \centering
    \includegraphics[width=0.65\textwidth]{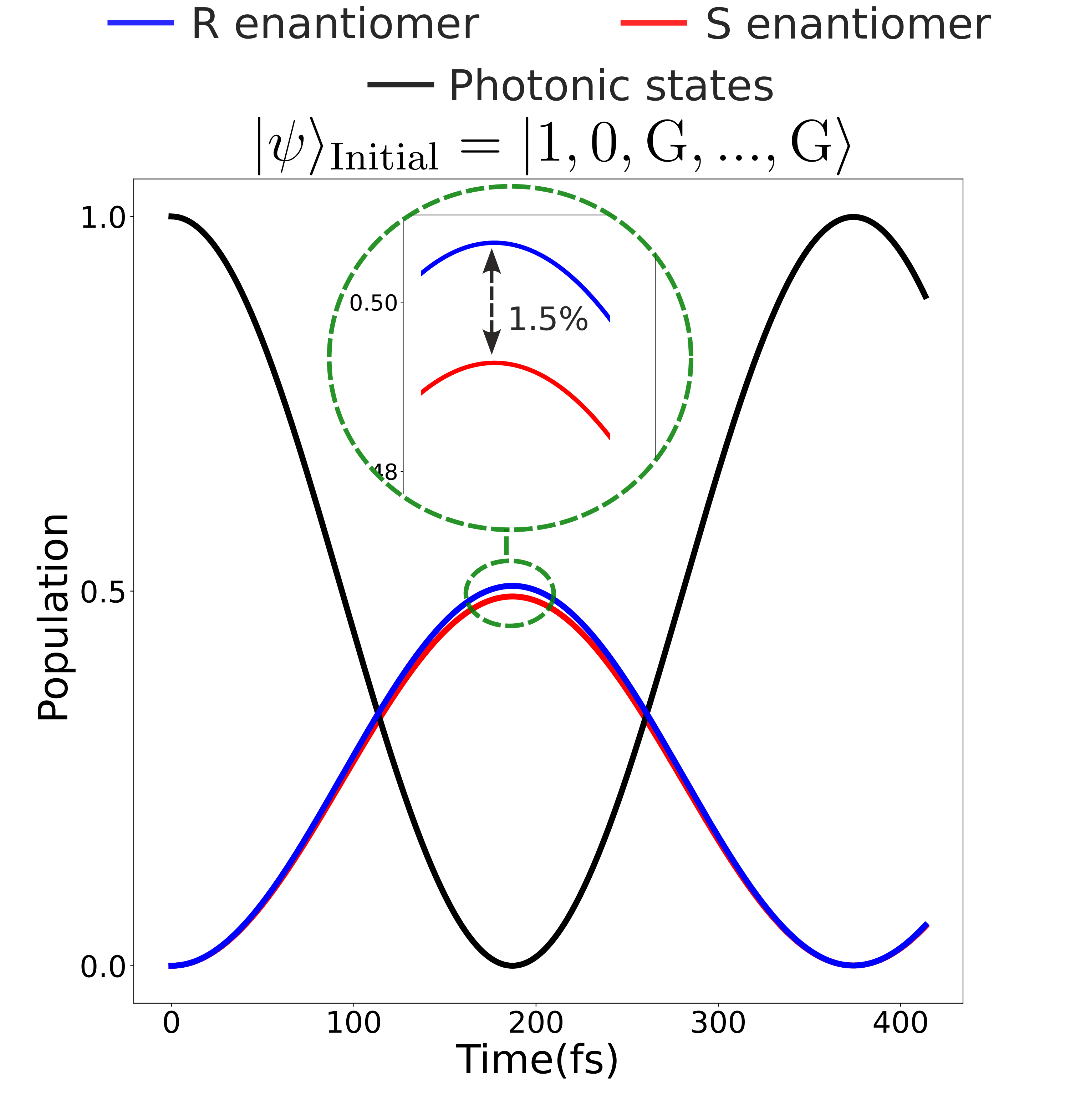}
    \caption{Time evolution of a photonic excitation in the cavity. Since the polaritons are asymmetric the tune propagation leads to an asymmetric energy redistribution.}
    \label{fig:time_prop}
\end{figure}
\begin{figure}
    \centering
    \includegraphics[width=0.65\textwidth]{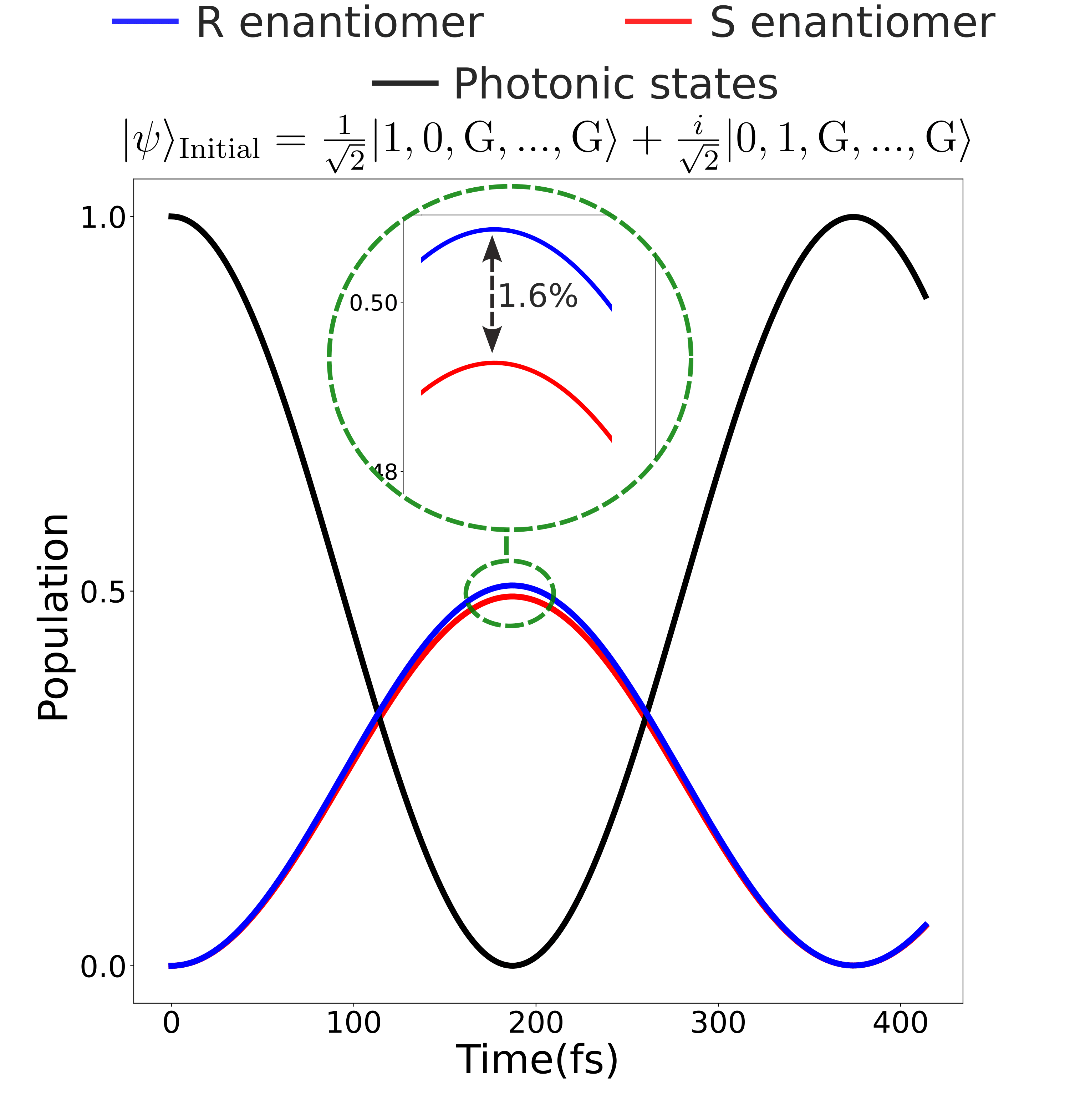}
    \caption{Time evolution of a photonic excitation in the cavity for a different initial photonic state.}
    \label{fig:second_time}
\end{figure}
\noindent P-tolyl sulfoxide is a chiral system that undergoes enantiomeric interconversion through photochemical activation. A schematic drawing of its $S_0$ and $S_1$ potential energy surfaces (PES) is reported in Fig.\ref{fig:enter-label}. Since the electronic Hamiltonian does not contain any parity-violating term, the surface is symmetric before and after the transition geometry, which is not chiral. Shining light through the mirrors is a viable way to prepare the strongly coupled system in a photonic initial state
\begin{equation}
 \ket{\psi}_{\textrm{Initial}}=C_{\alpha}\ket{1,0,G,...,G}+e^{i\theta}\sqrt{1-C^{2}_{\alpha}}\ket{0,1,G,...,G}.  
\end{equation}
The time evolution of $\ket{\psi}_{\textrm{Initial}}$ allows us to access the polaritons. Since the polaritons are asymmetric in the population of the enantiomers, the field excitation is distributed asymmetrically between the two configurations. In Fig.\ref{fig:time_prop}, we show the redistribution of energy between light and matter in a LHCP cavity filled with a racemic mixture of 3720 P-tolyl sulfoxides. The transition moments are computed at the CCSD level using the aug-cc-pVDZ basis set. The coupling parameter is equal to 0.0002 a.u. . %We illustrate how the excitation is reallocated between photonic states R and S enantiomers as a function of the propagation time in a LHCP cavity.
The initial state for the simulation in Fig.\ref{fig:time_prop} is 
\begin{equation}
 \ket{\psi}_{\textrm{Initial}}=\ket{1,0,G,...,G} \hspace{2cm} (C_{\alpha}=1),      
\end{equation}
and the time evolution is obtained by propagating the time-dependent Schrodinger equation:
\begin{equation}
\ket{\psi(t)} = e^{iH_{JC}t}\ket{\psi}_{\textrm{Initial}}.    
\end{equation}
We notice that when the excitation energy is transferred from the cavity photons to the p-tolyl sulfoxide, the R enantiomers host up to 1.5$\%$ more energy than the S enantiomers. Since the molecule then undergoes photochemical interconversion, this means that the mixture naturally becomes richer in the non-preferred enantiomers (the S enantiomer in this case). Control of the discrimination is achieved by modulating the initial wave function. For example, in Fig.\ref{fig:second_time}, we report an increase in the discrimination when the cavity is prepared in the state 
\begin{equation}
\ket{\psi}_{\textrm{Initial}}=\frac{1}{\sqrt{2}}\ket{1,0,G,...,G}+\frac{i}{\sqrt{2}}\ket{0,1,G,...,G}.  
\end{equation} 
Additional effects like thermal dissipation of the excitation energy, as well as the yield of the photochemical reaction, can modify the outcome of the time evolution. However, these effects do not change the qualitative conclusions of this analysis.  
\section{Conclusion}
In this work, we demonstrated that strong coupling between circularly polarized fields and chiral molecules effectively differentiates the photochemistry of different enantiomers. Due to the circular dichroism, Rabi splittings inside chiral cavities are specific to the configuration of the selected chiral molecule. While orientational effects significantly influence cavity phenomena, averaging over all configurations does not eliminate enantiospecific signatures in the Rabi splittings. The discrimination of the cavity exhibits collective behavior typical of strong coupling phenomena, increasing when the number of coupled molecules increases. In addition, we observe an asymmetry in the excitation distribution of the polaritonic states. These states are enriched in the enantiomer that are more strongly coupled to the field, even after orientational averaging. This result illustrates a field-induced condensation of the excitation on a single type of enantiomer, potentially leading to selective photochemical activation. Access to the polaritonic states can be achieved by illuminating the cavity through the mirrors. The desired asymmetry can be further controlled by altering the initial state of the cavity field, as demonstrated in Figure \ref{fig:time_prop}.
Our results are general and should also apply to chiral molecules in strong coupling with circularly polarized fields in the vibrational energy range. This could lead to enantiospecific cavity-induced catalysis or deceleration, in line with the results reported in Refs.\citenum{thomas2016ground,thomas2019tilting,hutchison2012modifying,ahn2023modification}. This topic will be the focus of a subsequent study. Future investigations will also deal with the exploration of tools to enhance the field-induced discrimination effects, such as the use of magnetic fields, in line with the enhancements observed in electronic and vibrational circular dichroism spectroscopies.
In conclusion, our findings, in conjunction with other works\cite{riso2023strongg,riso2023strong,baranov2022towards,schafer2023chiral,vu2022enhanced,mauro2022charge}, provide the necessary motivation for further developments in the use of strong light-matter coupling for enantiomeric separation. This research opens new pathways to unlock selective photochemical processes.
\section{Data availability} 
The geometries used for the calculations reported in this work can be found on the Zenodo open repository \cite{Zenodo}. Additional results on the effects of a finite $\mathbf{k}$ dimensionality of the sample in the cavity can be found in the Supplementary Material. Tests on the effects of the dipole approximation for the field are also included. 
\section{Code availability} 
The eT program\cite{folkestad20201} is open source and installation instructions can be found in the Zenodo repository \cite{Zenodo}.
\section{Acknowledgements}
R.R.R. thanks Carlos Bustamante, Dominik Sidler, Mark K. Svendsen, Christian Sch{\"a}fer and Michael Ruggenthaler for insightful discussions. R. R. R, M.C. and H. K. acknowledge funding from the Research Council of Norway through FRINATEK Project No. 275506. This work has received funding from the European Research Council (ERC) under the European Union’s Horizon 2020 Research and Innovation Programme (Grant Agreement No. 101020016). E. R. acknowledges funding from the European Research Council (ERC) under the European Union’s Horizon Europe Research and Innovation Programme (Grant No. ERC-StG-2021-101040197—QED-Spin).
\bibliography{bibliography}
\clearpage


\section*{Supplementary material}

# Summplementary Material to "Chiral polaritonics: cavity-mediated enantioselective excitation condensation"

Rosario R. Riso[1*], Matteo Castagnola[1], Enrico Ronca[2], Henrik Koch[1]

[1]Department of Chemistry, Norwegian University of Science and Technology, 7491 Trondheim, Norway

[2]Dipartimento di Chimica, Biologia e Biotecnologie, Università degli Studi di Perugia, Via Elce di Sotto, 8, 06123, Perugia, Italy

E-mail: `henrik.koch@ntnu.no, rosario.r.riso@ntnu.no`

## 1. Three dimensional sample

In Figs.8-9 of the main paper we showed that displacing two molecules 100Å apart from each other led to different effects depending on whether the displacement was parallel or perpendicular to the field wave vector. In Fig.1 we illustrate the effects that a finite 15nm z-dimension has on the properties of the system described in Fig.10. No significant variation of the effects is observed.

## 2. Effect of the dipole approximation

When the field wavelength is much larger than the molecular dimension the dipole approximation can be invoked. Within this framework, the exponential factor $e^{ikr}$ is approximated to 1 over the molecular dimension, i.e. the field does not vary over the molecular dimension. In Ref. 1, we showed that, for ground state properties, retaining the exponential factor in the light-matter interaction is critical to observe field-induced enantiomeric discrimination. In Fig.2, we show that the same statement is true also for excited state properties. If the dipole approximation is adopted, indeed, we see that the Rabi splitting is not dependent on the enantiomeric purity of the system. Moreover, the difference in the polaritonic population follows exactly the sample's stoichiometry.

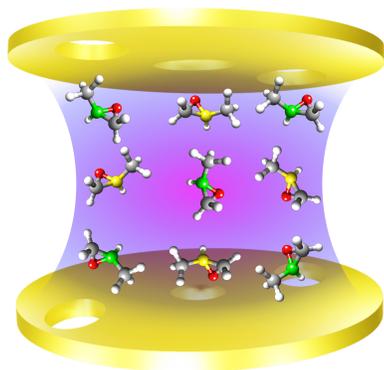
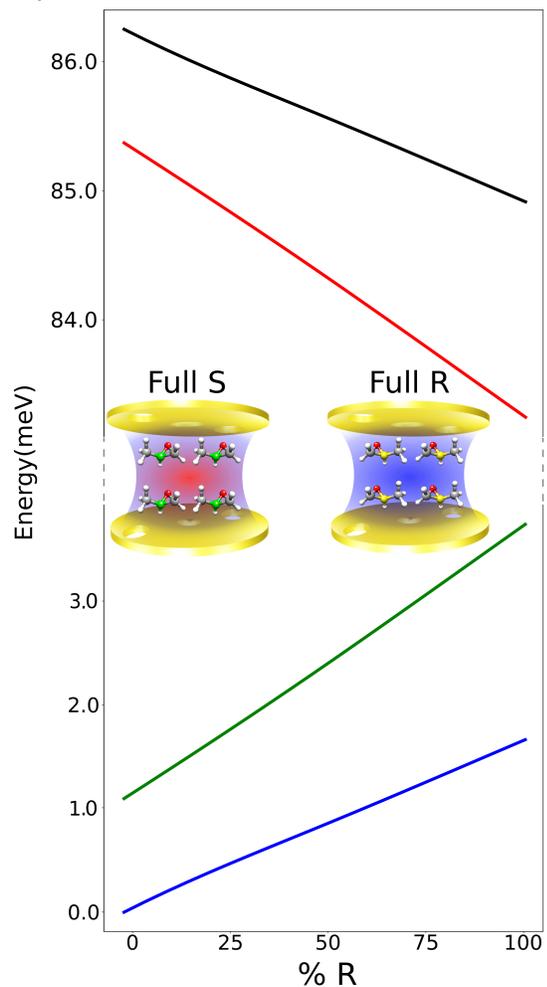
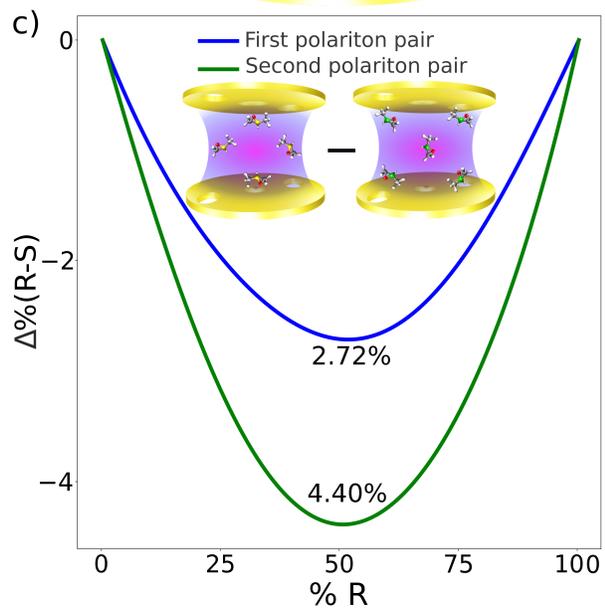

**Figure 1.** a) Collective behavior of a 3D sample of methyloxirane molecules in a chiral cavity. b) Excitation energies of the system as a function of the enantiomeric purity. c) Population difference between R and S enantiomers in the polaritonic states and in the enantiomeric mixture. The effects are very similar to the ones reported in Fig.10.

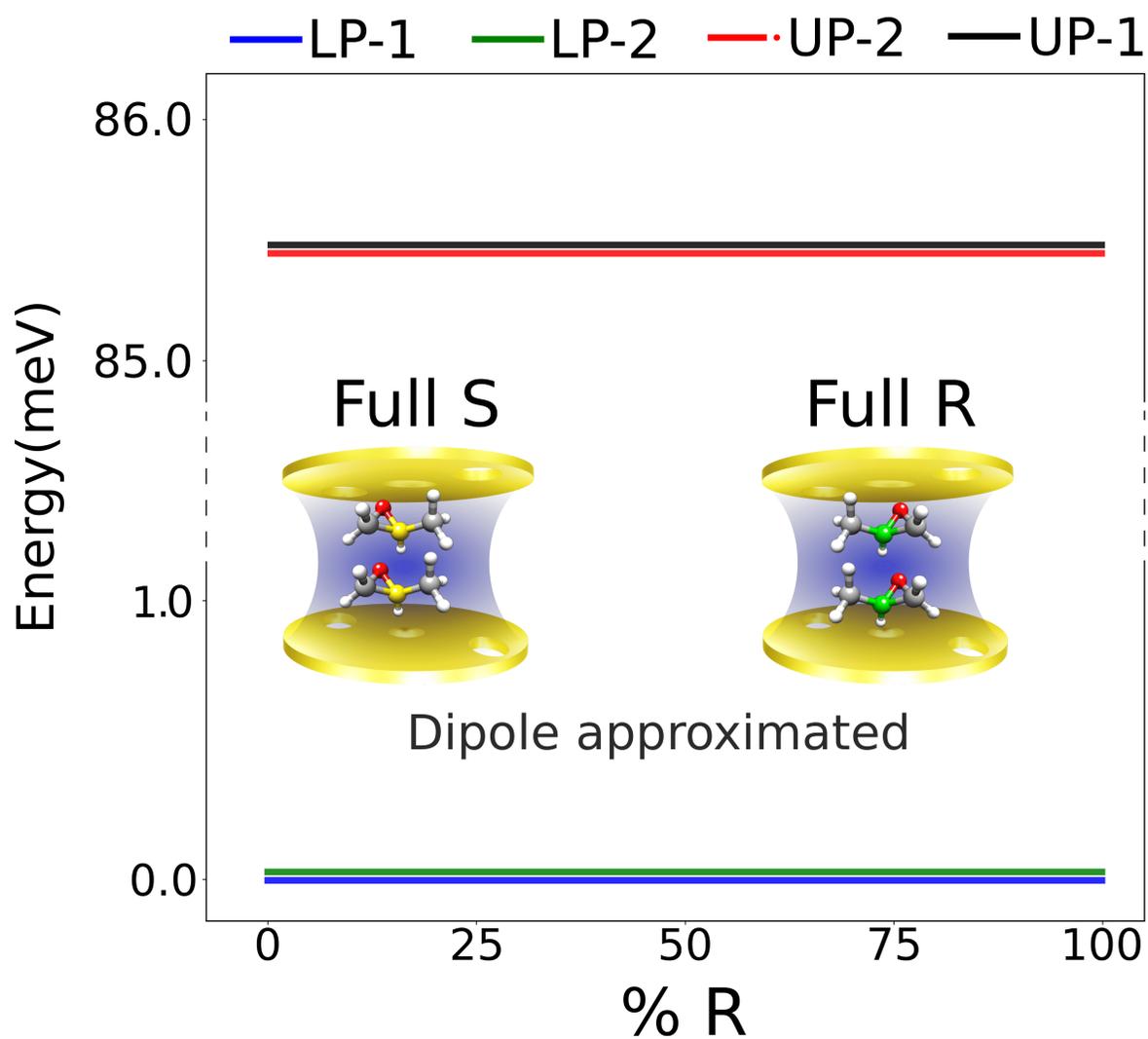

**Figure 2.** No change in the polaritonic states is observed when the enantiomeric composition is varied if the dipole approximation is adopted.

# TOC Graphic

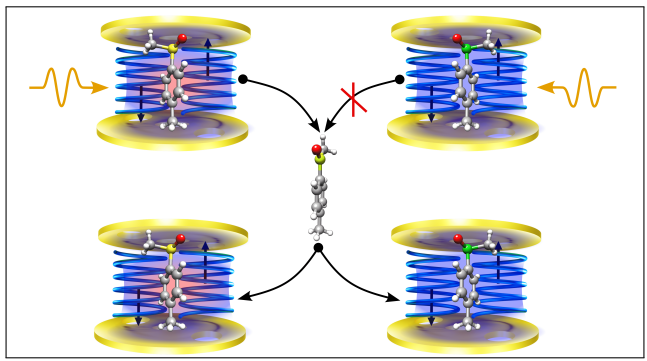


\end{document}